\documentclass[superscriptaddress,aps,preprintnumbers,nofootinbib]{revtex4-2}
\usepackage{amsmath,amssymb,amsfonts,graphicx,subfigure,slashed,amsthm,mathtools,upgreek,enumerate,tensor}
\usepackage{color}
\definecolor{darkblue}{rgb}{0.15,0.35,0.55}
\definecolor{reddish}{rgb}{0.65, 0.2, 0.2}
\definecolor{phthaloblue}{rgb}{0.0, 0.06, 0.54}
\definecolor{purple}{rgb}{0.5 ,0, 0.7}
\definecolor{bluegreen}{rgb}{0, 0.45, 0.35}
\definecolor{sakura}{rgb}{1 ,0.52, 0.74}
\definecolor{wakakusa}{rgb}{0.45 ,0.74, 0}
\definecolor{brown}{rgb}{0.48 ,0.23, 0}
\definecolor{skyblue}{rgb}{0.21 ,0.7, 1.}
\definecolor{purplegray}{rgb}{0.35,0.35,0.73}
\usepackage{braket}
\usepackage{autobreak}
\usepackage[bookmarks,linktocpage, colorlinks=true, plainpages = false, citecolor = darkblue,  linkcolor=darkblue, urlcolor = darkblue, filecolor = darkblue]{hyperref}

\usepackage{acro}
\DeclareAcronym{gr}{
    short=GR ,
    long=general relativity
}
\DeclareAcronym{uv}{
    short=UV ,
    long=ultraviolet
}
\DeclareAcronym{ir}{
    short=IR ,
    long=infrared
}
\DeclareAcronym{wdw}{
    short=WDW ,
    long=Wheeler-DeWitt
}
\DeclareAcronym{hl}{
    short=HL ,
    long=Ho\v{r}ava-Lifshitz
}
\DeclareAcronym{rg}{
    short=RG ,
    long=renormalization group
}
\DeclareAcronym{adm}{
    short=ADM ,
    long={Arnowitt, Deser and Misner}
}

\begin{document}
\preprint{YITP-23-123,  IPMU23-0035}

\title{
Hartle-Hawking No-boundary Proposal and Ho\v{r}ava-Lifshitz Gravity}

\author{Hiroki Matsui}
\email{hiroki.matsui@yukawa.kyoto-u.ac.jp}
\affiliation{Center for Gravitational Physics and Quantum Information, Yukawa Institute for Theoretical Physics, Kyoto University, Kitashirakawa Oiwakecho, Sakyo-ku, Kyoto 606-8502, Japan}
\author{Shinji Mukohyama}
\email{shinji.mukohyama@yukawa.kyoto-u.ac.jp}
\affiliation{Center for Gravitational Physics and Quantum Information, Yukawa Institute for Theoretical Physics, Kyoto University, Kitashirakawa Oiwakecho, Sakyo-ku, Kyoto 606-8502, Japan}
\affiliation{Kavli Institute for the Physics and Mathematics of the Universe (WPI), The University of Tokyo Institutes for Advanced Study, The University of Tokyo, Kashiwa, Chiba 277-8583, Japan}

\begin{abstract}
We study the Hartle-Hawking no-boundary proposal in the framework of Ho\v{r}ava-Lifshitz gravity. The former is a prominent hypothesis that describes the quantum creation of the universe, while the latter is a potential theory of quantum gravity that ensures renormalizability and unitarity, at least in the so-called projectable version. For simplicity, we focus on a global universe composed of a set of local universes each of which is closed, homogeneous and isotropic. Although applying the no-boundary proposal to Ho\v{r}ava-Lifshitz gravity is not straightforward, we demonstrate that the proposal can be formulated within the Ho\v{r}ava-Lifshitz gravity utilizing the Lorentzian path integral formulation of quantum gravity. In projectable Ho\v{r}ava-Lifshitz gravity, the no-boundary wave function of the global universe inevitably contains entanglement between different local universes induced by ``dark matter as integration constant''. On the other hand, in the non-projectable version, the no-boundary wave function of the global universe is simply the direct product of wave functions of each local universe. We then discuss how the no-boundary wave function is formulated under Dirichlet and Robin boundary conditions. For the Dirichlet boundary condition, we point out that its on-shell action diverges due to higher-dimensional operators, but this problem can in principle be ameliorated by taking into account the renormalization group flow. However, utilizing the Picard-Lefschetz theory to identify the relevant critical points and performing the complex lapse integration, we find that only the tunneling wave function can be obtained, as in the case of general relativity. On the other hand, for the Robin boundary condition with a particular imaginary Hubble expansion rate at the initial hypersurface, the no-boundary wave function can be achieved in the Ho\v{r}ava-Lifshitz gravity. 
\end{abstract}
\date{\today}
\maketitle

\section{Introduction}

Given that quantum theory is universally applicable, the entire universe is also regarded as requiring a quantum approach. For late-time cosmology, the classical theory may be appropriate because the universe is sufficiently large and the expansion rate is sufficiently low. However, in the early days of the universe, quantum effects are believed to have played a central role. Considering the quantum fluctuations inherent to the universe, it is plausible that the universe originated from nothing devoid of any space-time. This idea is a cornerstone of quantum cosmology, with a long history dating back to Lemaitre~\cite{Lemaitre:1931zzb}. The most robust formulations of this idea, such as the no-boundary proposal~\cite{Hartle:1983ai}, the tunneling proposal~\cite{Vilenkin:1984wp}, and the DeWitt’s proposal~\cite{DeWitt:1967yk} are given by imposing certain boundary conditions on the wave functional of quantum gravity, called the wave function of the universe. In quantum cosmology, a conventional approach to describing the wave function of the universe utilizes the path integral of quantum gravity over specific $4$-dimensional geometry, represented as $\Psi[g]=\int\mathcal{D} g^{(4)}\, e^{i S[g^{(4)}]/\hbar}$. Here, the $4$-dimensional metric $g^{(4)}$ is restricted to those inducing the spatial metric $g$ on the $3$-geometry, and the diffeomorphism invariance is properly treated. Conversely, an alternative approach to formulating the wave function of the universe involves the canonical quantization of quantum gravity, and seeks solutions to the \ac{wdw} equation, $\hat{H}[g]\Psi[g]=0$ with boundary conditions where $\hat{H}[g]$ denotes the Hamiltonian operator.

Historically, the Hartle-Hawking no-boundary proposal postulated that the wave function of the universe is characterized by a path integral over all compact Euclidean geometries that possess solely a $3$-dimensional boundary~\cite{Hartle:1983ai}. However, in more contemporary interpretations, this path integral is evaluated from a $3$-geometry that is essentially of null volume or nothing, to a finite $3$-geometry~\cite{Halliwell:1988ik}. Notably, recent progress has been achieved in the rigorous analyses of the Lorentzian path integral for both the no-boundary and tunneling proposals, elucidating that these are in fact identical~\cite{Feldbrugge:2017kzv} and perturbations around the background geometry are unstable~\cite{Feldbrugge:2017fcc,Feldbrugge:2017mbc}. Although there have been numerous studies and discussions on this issue~\cite{Feldbrugge:2017fcc, DiazDorronsoro:2017hti, Feldbrugge:2017mbc, Feldbrugge:2018gin, DiazDorronsoro:2018wro,Halliwell:2018ejl,Janssen:2019sex, Vilenkin:2018dch, Vilenkin:2018oja, Bojowald:2018gdt, DiTucci:2018fdg, DiTucci:2019dji, DiTucci:2019bui, Lehners:2021jmv,Matsui:2021yte,Martens:2022dtd,Matsui:2022lfj}, the only solution that seems plausible is to change the boundary conditions or the boundary terms of the background geometry at the price of abandoning the Dirichlet boundary condition, \textit{i.e}. the notion of a sum over compact and regular geometries in the no-boundary proposal~\cite{DiTucci:2019dji,DiTucci:2019bui}~\footnote{In Refs.~\cite{Vilenkin:2018dch,Vilenkin:2018oja} the authors proposed a boundary term for the gravitational action of the linearized perturbations to satisfy the Robin boundary condition.}. Consequently, it is considered that alterations to the framework of the no-boundary proposal, such as the modification of boundary conditions or the boundary terms of the background geometry, are inevitable. There has been some work on Lorentzian quantum cosmology based on these different boundary conditions~\cite{DiTucci:2019dji,DiTucci:2019bui,Narain:2021bff,
Narain:2022msz,Ailiga:2023wzl}.

On the other hand, these consequences of quantum cosmology are usually discussed only in the framework of \ac{gr} and may need to be verified in the framework of the \ac{uv} completion of quantum gravity. As well known, the pursuit of the quantum field theory of gravity has encountered numerous challenges. One of the most enduring issues is the non-renormalizability of \ac{gr}. This non-renormalizability induces uncontrolled \ac{uv} divergences, causing the theory to break down perturbatively. Although it is possible to achieve renormalizability~\cite{Stelle:1976gc}, or even super-renormalizability~\cite{Asorey:1996hz}, by introducing higher curvature operators into the Einstein-Hilbert action, this approach unfortunately introduces massive ghosts. These ghosts manifest as a non-unitary quantum theory in the \ac{uv} domain. Consequently, the challenge of reconciling renormalizability and unitarity has consistently impeded the construction of a consistent theory of quantum gravity.

\ac{hl} gravity~\cite{Horava:2009uw} is a theory of quantum gravity that aims to harmonize the renormalizability with unitarity beyond \ac{gr}. The theory is based on the so-called \textit{anisotropic scaling}, or \textit{Lifshitz scaling} between time and space coordinates,
\begin{equation}
 t \to b^z t\,, \quad \vec{x} \to b \vec{x}\,, 
\end{equation}
where $t$ is the time coordinate, $\vec{x}$ represents the spatial coordinates vector and $z$ is a number called \textit{dynamical critical exponent}. This results in the inclusion of only higher-order spatial curvature operators making the theory power-counting renormalizable and the equations of motion in \ac{hl} gravity are restricted to up to second-order time derivatives, thus avoiding the presence of Ostrogradsky ghosts. The anisotropic scaling in the \ac{uv} regime in $3+1$ dimensions is with $z=3$, which breaks Lorentz symmetry but ensures renormalizability. In the \ac{ir} regime, the usual scaling of $z=1$ is recovered. Additionally, an anisotropic scaling of $z=3$ also provides a mechanism for generating scale-invariant cosmological perturbations, thereby solving the horizon problem~\cite{Mukohyama:2009gg} without the need for inflation and offering a possible solution to the flatness problem~\cite{Bramberger:2017tid}.

In this paper, we discuss the Hartle-Hawking no-boundary proposal in the framework of the \ac{hl} gravity. As already mentioned, the \ac{hl} gravity provides a picture of the early universe on the basis of the renormalizability of quantum gravity~\cite{Mukohyama:2009gg,Bramberger:2017tid} and in the framework of quantum cosmology, this theory liberates the DeWitt's proposal~\cite{DeWitt:1967yk}~\footnote{Formulating the wave function based on the DeWitt's proposal using path integral formulation is not straightforward. For a formulation of DeWitt's proposal based on the path integral, see Ref.~\cite{Matsui:2023tkw}.}, which suggests that the wave function of the universe should vanish at the classical big-bang singularity, from the perturbation problem~\cite{Matsui:2021yte,Martens:2022dtd}. On the other hand, however, it is found that application of the Hartle-Hawking no-boundary proposal to the \ac{hl} gravity is not straightforward~\cite{Bertolami:2011ka,Bramberger:2017tid}, and for instance, the Euclidean on-shell action diverges due to the higher-order curvature terms in \ac{hl} gravity, indicating the breakdown of the semi-classical approximation for the no-boundary wave function. This problem also exists when we define the wave function using the \ac{wdw} equation. Around the initial singularity, the super-potential diverges due to the higher-order curvature terms, making it suitable to take DeWitt's boundary condition~\cite{DeWitt:1967yk} instead of the no-boundary proposal. However, we show that the modern framework of the no-boundary proposal, utilizing the Lorentzian path integral formulation of quantum gravity with the modification of boundary conditions or the boundary terms of the background manifold~\cite{DiTucci:2019dji,DiTucci:2019bui} liberates the above issue and allows for a systematic analysis of the no-boundary wave function based on the \ac{hl} gravity. We also discuss the possibility that the divergence of the on-shell action due to higher-dimensional operators may be ameliorated by taking into account the \ac{rg} flow.

We consider the projectable version of \ac{hl} gravity, which has been definitively demonstrated to be perturbatively renormalizable~\cite{Barvinsky:2015kil,Barvinsky:2017zlx}. While the \ac{hl} gravity is power-counting renormalizable, the projectable version extends this property to be absolutely and perturbatively renormalizable, without any theoretical inconsistencies.
Also, the projectable theory is phenomenologically viable as a theory
of gravity if a dimensionless coupling constant, usually denoted as
$\lambda$, sufficiently quickly flows to $1$ in the \ac{ir} under the \ac{rg}
flow so that the \ac{ir} instability of the scalar graviton does not show
up~\cite{Mukohyama:2010xz} and that an analogue of the Vainshtein
mechanism takes
place~\cite{Mukohyama:2010xz,Izumi:2011eh,Gumrukcuoglu:2011ef}. On the
other hand, the recovery of Lorentz invariance in the matter sector
remains an open issue in Ho\v{r}ava-Lifshitz gravity in general (see
e.g. \cite{Coates:2018vub}).
Besides, as we will elaborate on this, the projectable \ac{hl} gravity does not have the local Hamiltonian constraint but imposes the global Hamiltonian constraint on the system. Thus, the Lorentzian path integral approach in projectable \ac{hl} gravity is non-trivial and we first discuss the need to consider the Lorentzian path integral over a global universe composed of a set of local universes in projectable \ac{hl} gravity.

Then, we investigate the \ac{hl} wave function of the global universe, comprised of a set of local universes under various boundary conditions, focusing especially on Dirichlet and Robin boundary conditions. For the Dirichlet boundary condition, the Lorentzian path integral formulation gives the \ac{hl} wave function describing the quantum creation of the universe from nothing although its on-shell action diverges due to higher-dimensional operators. We comment that this problem can in principle be ameliorated by taking into account the \ac{rg} flow. However, even if avoiding the on-shell divergences, employing the Picard-Lefschetz theory to identify the relevant critical points and performing the integration over the complex lapse, we find that only the tunneling wave function can be obtained, as in the case of \ac{gr}. On the other hand, for the Robin boundary condition, the \ac{hl} wave function is consistently given if the Hubble expansion rate is finite on the initial hypersurface. In particular, if the initial Hubble expansion rate is finite and imaginary, the no-boundary wave function can be obtained. However, we note that the analysis of the \ac{hl} wave function using the saddle-point method based on the Picard-Lefschetz theory becomes technically challenging in this case.

The rest of the present paper is organized as follows. In Section~\ref{sec:HL-graivty}, we provide a brief review of the construction of the projectable \ac{hl} gravity in $3 + 1$ dimensions. In Section~\ref{sec:Lorentzian-path-integral}, we review the Hartle-Hawking no-boundary proposal based on the Lorentzian path integral formulation and discuss its application to the projectable \ac{hl} gravity. In Section~\ref{sec:no-boundary-proposal}, we discuss the Hartle-Hawking no-boundary proposal involving various boundary conditions and related terms for the background spacetime manifold. In particular, we focus on the Dirichlet and Robin boundary conditions and discuss how the no-boundary wave function is realized in the framework of projectable \ac{hl} gravity. In Section~\ref{sec:discussions-conclusions} we conclude our work.

\section{Projectable Ho\v{r}ava-Lifshitz gravity}
\label{sec:HL-graivty}

In this section, we briefly introduce the basic framework of \ac{hl} gravity. The basic variables are the lapse function $N$, the shift vector $N^i$ and the spatial metric $g_{ij}$ with the positive definite signature $(+,+,+)$. In the \ac{ir}, one can construct the $3 + 1$ dimensional metric out of the basic variables as in the \ac{adm} formalism~\cite{Arnowitt:1962hi}, 
\begin{align}
\mathrm{d} s^2 = N^2 \mathrm{d} t^2 + g_{i j} (\mathrm{d} x^i + N^i \mathrm{d} t) (\mathrm{d} x^j + N^j \mathrm{d} t)\,,
\end{align}
The shift vector $N^i$ and the spatial metric $g_{ij}$ in general depend on all four coordinates. On the other hand, we assume that the lapse function $N$ is a function of time only, i.e. $N=N(t)$, in the projectable \ac{hl} gravity. While a naive non-projectable extension would lead to phenomenological obstacles~\cite{Charmousis:2009tc} and theoretical inconsistencies~\cite{Li:2009bg}, one can obtain a classically consistent non-projectable version by inclusion of terms depending on the spatial derivatives of the lapse function in the action. However, consistency at the quantum level, such as renormalizability, 
\footnote{The strategy of~\cite{Barvinsky:2015kil} can not be applied
to the non-projectable version. However, this does not necessarily
mean that the non-projectable theory is non-renormalizable. Therefore
it is still an open question whether the non-projectable theory is
renormalizable or not.}
has not been established for the non-projectable version. In the present paper, we mainly focus on the projectable \ac{hl} gravity, but our main results can extend to the non-projectable version as we shall discuss later.

The $3 + 1$ dimensional action $S$ describing projectable \ac{hl} gravity is written by~\cite{Mukohyama:2010xz},
\begin{equation}
\label{action}
S= \frac{{\cal M}_{\rm HL}^{2}}{2}\int \mathrm{d} t \mathrm{d}^3 x N \sqrt{g} \left( K^{ij}K_{ij}-\lambda K^2 +c^2_gR-2\Lambda+{\mathcal O}_{z>1} \right)\,, 
\end{equation}
where ${\cal M}_{\rm HL}$ is the overall mass scale. The extrinsic curvature tensor $K_{ij}$ is defined by $K_{ij}= (\partial_t g_{ij}- g_{j k}\nabla_iN^k-g_{i k}\nabla_jN^k)/(2N)$, with $\nabla_i$ being the spatial covariant derivative compatible with $g_{ij}$, $K^{ij}=g^{ik}g^{jl}K_{kl}$, $K=g^{ij}K_{ij}$ and $R$ is the Ricci scalar of $g_{ij}$, where $g^{ij}$ is the inverse of $g_{ij}$. The constants $\Lambda$ and $c_g$ are the cosmological constant and the propagation speed of tensor gravitational waves. The higher dimensional operators ${\mathcal O}_{z>1}$ is given by
\begin{align}
\frac{{\mathcal O}_{z>1}}{2} & = 
c_1\nabla_iR_{jk}\nabla^iR^{jk}+c_2\nabla_iR\nabla^iR+c_3R_i^jR_j^kR_k^i\nonumber\\
 &  +c_4RR_i^jR_j^i + c_5R^3 
  + c_6R_i^jR_j^i+c_7R^2 \,.
\end{align}
All coupling constants in the action, $\Lambda$ and $c_g$ mentioned above as well as $\lambda$ and $c_{n}$ ($n=1,\cdots,7$) are subject to running under the \ac{rg} flow. In the \ac{uv} regime, terms with two time derivatives and those with six spatial derivatives are dominant. Conversely, in the infrared (IR) regime, the higher derivative terms become less significant, leading the theory to the standard scaling with $z=1$. Furthermore, as the parameter $\lambda$ approaches unity in the IR limit, and if this convergence is fast enough, the theory reverts to \ac{gr} (plus built-in dark matter) via a mechanism analogous to the Vainshtein mechanism~\cite{Mukohyama:2010xz,Izumi:2011eh,Gumrukcuoglu:2011ef}. Also, the linear \ac{ir} instability associated with the scalar graviton does not show up under a certain condition~\cite{Mukohyama:2010xz}.

We assume the $3$-dimensional space of the model to be the union of connected pieces $\Sigma_{\alpha}$ ($\alpha=1,\dots$), each of which we call a \textit{local $3$-dimensional universe}. The union of all $\Sigma_{\alpha}$ would thus represent the entire universe. In such a configuration, we have a set of shift vectors $N^i = N_{\alpha}^i(t,x)$ and spatial metrics $g_{ij} = g^{\alpha}_{ij}(t,x)$, which are different for different local universes $\Sigma_{\alpha}$, while the lapse function $N=N(t)$ is common to all pieces. The variation of the projectable \ac{hl} action with respect to $N$ yields the \textit{global Hamiltonian constraint} 
\begin{equation}
    \label{eq:Hamiltonian-constraint}
    \sum_{\alpha} \int_{\Sigma_{\alpha}}\mathrm{d}^3x\, \mathcal{H}_{g\perp}=0 \, ,
    \quad \text{with} \quad 
    \mathcal{H}_{g\perp} = \frac{{\cal M}_{\rm HL}^{2}}{2}\sqrt{g}( K^{ij}K_{ij}-\lambda K^2 - c^2_gR + 2\Lambda -{\mathcal O}_{z>1} )\,.
\end{equation}
While this constraint sets the sum of contributions from all $\Sigma_{\alpha}$ to vanish, each contribution does not have to vanish, \textit{i.e}.
\begin{equation}
    \int_{\{\Sigma_{\alpha}\}}d^3x\, \mathcal{H}_{g\perp} \ne 0 \, .
\end{equation}
Therefore, if we are interested in one local universe, \textit{i.e}. an element of $\{\Sigma_{\alpha}\}$, the \textit{Hamiltonian constraint} does not need to be enforced. This results in ``dark matter as integration constant''~\cite{Mukohyama:2009mz,Mukohyama:2009tp}. On the other hand, the non-projectable version imposes the local Hamiltonian constraint at each spatial point in each local universe. As a result, the non-projectable theory does not allow ``dark matter as integration constant'', contrary to the projectable theory.

\section{Lorentzian path integral for projectable Ho\v{r}ava-Lifshitz gravity}
\label{sec:Lorentzian-path-integral}

In this section, we will introduce the wave function of the universe based on the Lorentzian path integral. Originally, in the Hartle-Hawking no-boundary proposal, the wave function of the universe is suggested to be given by a path integral over all compact Euclidean geometries that have $3$-dimensional geometry configuration as the only boundary. This elegantly explains the quantum birth of the universe in principle but is considered to be certainly incomplete for various technical reasons. For instance, the conformal factor problem arises due to the fact that the Euclidean action of gravity is not bounded from below and suggests that the Euclidean path integral diverges and is not well-defined~\cite{Gibbons:1978ac}. Recently, the no-boundary and tunneling proposals in minisuperspace quantum cosmology have been investigated by the Lorentzian path integral formulation~\cite{Feldbrugge:2017kzv,DiazDorronsoro:2017hti}. Integrals of phase factors such as $e^{i S_{\rm GR}/\hbar}$ usually do not manifestly converge, but the convergence can be achieved by shifting the contour of the integral onto the complex plane by applying Picard-Lefschetz theory~\cite{Witten:2010cx}. According to Cauchy's theorem, if the integral does not have poles in a region on the complex plane, the Lorentzian nature of the integral is preserved even if the integration contour on the complex plane is deformed within such a region. In particular, as we will show later, the path integral can be rewritten as an integral over the gauge-fixed lapse function $N$, and directly performed, unlike the Euclidean path integral formulation.

Hereafter, based on the previous discussions about projectable \ac{hl} gravity in which the local Hamiltonian constraint is not imposed, but the global Hamiltonian constraint is enforced, we make two assumptions. First, we assume that each connected space $\Sigma_{\alpha}$ is a closed universe. Second, for simplicity, we assume that each closed universe is described by a closed Friedmann-Lema\^{i}tre-Robertson-Walker (FLRW) metric, 
\begin{equation}
    N_{\alpha}^i = 0\,, \quad g^{\alpha}_{ij} = a^2_{\alpha}(t) \left[\Omega_{ij} ({\bf x}) \right]\,,
\end{equation}
where $\Omega_{ij}$ is the metric of the unit $3$-sphere with the curvature constant set to $1$, \textit{i.e}. the Riemann curvature of $\Omega_{ij}$ is simply $\delta^i_k\delta^j_l-\delta^i_l\delta^j_k$. Given this definition, the spatial indices $i, j, \dots$ are thus raised and lowered by $\Omega^{ij}$ and $\Omega_{ij}$ respectively.

We shall start the projectable \ac{hl} action of the following form, 
\begin{align}
S\left[\{a_{\alpha}\},N\right] &=  \int_{t_i}^{t_f}\mathrm{d} t \sum_{\alpha}\, V\!\left(p_{a_{\alpha}}\dot{a}_{\alpha}- N \mathcal{H}(\{a_{\alpha},p_{a_{\alpha}}\}) \right)\,,
\end{align}
where $p_{a_{\alpha}}$ is the canonical momentum conjugate to the scale factor of each local universe, $a_{\alpha}$, and $V={\cal M}_{\rm HL}^{2}V_3$ in which $V_3=\int \mathrm{d}^3 x\sqrt{\Omega}$ is the volume of the unit $3$-sphere, $N$ represents a Lagrange multiplier which enforces the global Hamiltonian constraint, $\mathcal{H}(\{a_{\alpha},p_{a_{\alpha}}\})=0$. As previously mentioned, while the projectable \ac{hl} gravity does not ensure the local Hamiltonian constraint, it enforces the global Hamiltonian constraint. Consequently, the action of the global universe, which comprises several local universes, is formally analogous to the system of point particles in a parametrized form. As demonstrated in~\cite{Halliwell:1988wc} for \ac{gr}, the comprehensive method introduced by Batalin, Fradkin, and Vilkovisky can be applied to the projectable \ac{hl} gravity.

In the following discussion, we shall utilize the Batalin-Fradkin-Vilkovisky (BFV) formalism~\cite{Fradkin:1975cq,Batalin:1977pb} to construct the gravitational propagator preserving the global Hamiltonian constraint, and the BFV path integral reads~\cite{Halliwell:1988wc},
\begin{align}
G[\{a_{\alpha}\}]&\equiv \int \prod_{\alpha} \mathcal{D}a_{\alpha}\mathcal{D}p_{a_{\alpha}}\mathcal{D}\Pi\,\mathcal{D}N\,\mathcal{D}\rho\,\mathcal{D}\bar{c}\,\mathcal{D}\bar\rho\,\mathcal{D}c\,\exp(iS_{\rm BRS}/\hbar)\,,\\
S_{\rm BRS}&\equiv \int_{t_i}^{t_f} {\rm d}t 
\sum_{\alpha}V \left(p_{a_{\alpha}}\dot{a}_{\alpha}- N \mathcal{H} + \Pi \dot{N} +\bar\rho\dot{c} +\bar{c}\dot\rho-\bar\rho\rho\right)\,,
\end{align}
where $S_{\rm BRS}$ is the Becchi-Rouet-Stora (BRS) invariant action, including the Hamiltonian constraint $\mathcal{H}(\{a_{\alpha}\})$, a Lagrange multiplier $\Pi$ and ghost fields $\rho$, $\bar{\rho}$, $c$, $\bar{c}$, preserving the BRS symmetry, i.e. invariant under the following transformation,
\begin{align}
\begin{split}
\delta a_{\alpha}=\bar{\lambda} c\frac{\partial \mathcal{H} }{\partial p_{a_{\alpha}}}\,,\;\;\delta p_{a_{\alpha}}=-\bar{\lambda} c\frac{\partial \mathcal{H} }{\partial a_{\alpha}}\,,\;\;
\delta N= \bar{\lambda}\rho\,,\;\;
\delta\bar{c}=-\bar{\lambda}\Pi\,,\;\;\delta\bar\rho=-\bar{\lambda}\mathcal{H} \,,\;\; \delta \Pi=\delta c=\delta \rho=0\,,
\end{split}
\end{align}
where $\bar{\lambda}$ is a variable parameter. The ghost and multiplier parts can be integrated out, and eventually, we obtain,
\begin{align}\label{G-propagator}
G[\{a_{\alpha}\}] &= \int \! \mathrm{d}N(t_f-t_i)
\int \prod_{\alpha} \mathcal{D}a_{\alpha}\mathcal{D}p_{a_{\alpha}}\exp\left(i \int_{t_i}^{t_f} {\rm d}t \sum_{\alpha}V \left(p_{a_{\alpha}}\dot{a}_{\alpha} - N \mathcal{H}\right)/ \hbar\right)\nonumber
\\& = \int\! \mathrm{d}N(t_f-t_i) \int 
\prod_{\alpha} \mathcal{D}a_{\alpha} \exp\left(i S[\{a_{\alpha}\},N]
/ \hbar\right)\,,
\end{align}
which is the integral over $N(t_f-t_i)$ between the initial and final configurations.

With these assumptions the projectable \ac{hl} action reads, 
\begin{equation} \label{eq:HLaction-FLRW}
        S\left[\{a_{\alpha}\},N\right] =  \sum_{\alpha}\, V\!\int_{t_i}^{t_f} \mathrm{d} t\, 
        \left({Na_{\alpha}^3}\right) \left[
              \frac{3(1-3\lambda)}{2}\left(\frac{\dot{a}_{\alpha}}{Na_{\alpha}}\right)^{2}
              + \frac{\alpha_3}{a^6_{\alpha}}+\frac{\alpha_2}{a^4_{\alpha}}+c_{\rm g}^2
        \frac{3}{a^2_{\alpha}}-\Lambda \right]\,. 
\end{equation}
Here, $\alpha_i$ are constants that are linear combinations of coupling constants in the action. Redefining the variable $q_{\alpha}(t) \equiv \frac{2}{3}a_{\alpha}(t)^{3/2}$, and constants,
\begin{equation}
    \begin{gathered}
        g_K = 3c_{\rm g}^2\left(\frac{9}{4}\right)^{\frac{1}{3}}
 \frac{1}{3(3\lambda-1)} \,, \quad
            g_{\Lambda} = \Lambda\left(\frac{9}{4}\right)
             \frac{1}{3(3\lambda-1)} \,, \\
            g_2 = \alpha_{2}\left(\frac{4}{9}\right)^{\frac{1}{3}}
 \frac{1}{3(3\lambda-1)} \,,\quad
            g_3 = \alpha_{3}\left(\frac{4}{9}\right)
 \frac{1}{3(3\lambda-1)}\,,\quad  {\cal V}=3(3\lambda-1)V\,,
    \end{gathered}
\end{equation}
we obtain the following \ac{hl} action for the global universe, which comprises several local
universes, 
\begin{align}
S\left[\{q_{\alpha}\},N\right]
 &=\sum_{\alpha}\, {\cal V}\!\int_{t_i}^{t_f}Ndt \Biggl[-\frac{\dot{q}_{\alpha}^2}{2N^2}
 +g_Kq_{\alpha}^{\frac{2}{3}}
 -g_{\Lambda}q_{\alpha}^{2} +g_{2}q_{\alpha}^{-\frac{2}{3}}+g_{3}q_{\alpha}^{-2}\Biggr]\,. 
\end{align}

For the above action, the gravitational propagator can be written as,
\begin{equation}\label{G-propagator_0}
 G\left[\{q_{\alpha}(t_f)\};\{q_{\alpha}(t_i)\}\right] = \int \mathrm{d}N
 \int_{\{q_{\alpha}(t_i=0)\}}^{\{q_{\alpha}(t_f=1)\}}
\mathcal{D}q  \, e^{i S\left[\{q_{\alpha}\},N\right] / \hbar}\,,
 \end{equation}  
where we proceed with the lapse integral and the path integral over all configurations of the scale factor  $\{q_{\alpha}\}$ with the given initial $(t_i=0)$ and final $(t_f=1)$ values. To evaluate the above path integral we proceed with the semi-classical analysis, and utilize the classical solutions of the \ac{hl} action where we add possible boundary contributions $S_B$ localized on the hypersurfaces at $t_{i,f}=0,1$,
\begin{align}
S\left[\{q_{\alpha}\},N\right]&=\sum_{\alpha}
S\left[q_{\alpha},N\right]+S_B \notag  \\
 &=\sum_{\alpha}\, {\cal V}\!\int_{t_i=0}^{t_f=1} Ndt 
 \Biggl[-\frac{\dot{q}_{\alpha}^2}{2N^2}
 +g_Kq_{\alpha}^{\frac{2}{3}}
 -g_{\Lambda}q_{\alpha}^{2} +g_{2}q_{\alpha}^{-\frac{2}{3}}+g_{3}q_{\alpha}^{-2}\Biggr]
 +S_B\,. 
\end{align}
Hereafter, we proceed with the variation of the action and derive the equation of motion. Since the \ac{hl} action $S\left[q_{\alpha},N\right]$ depends on $q_{\alpha}$, $\dot{q}_{\alpha}$, the variation of the action is given by 
\begin{align}
\begin{split}
&\delta S\left[q_{\alpha},N\right] = {\cal V}\int_{t_i=0}^{t_f=1} dt \, \Bigl[
\frac{\partial S}{\partial q_{\alpha}}\delta q_{\alpha}+\frac{\partial S}{\partial \dot{q}_{\alpha}}\delta \dot{q}_{\alpha}\Bigr]
+ \delta S_B \\
&= {\cal V}\int_{t_i=0}^{t_f=1} Ndt \, \Bigl[\frac{\ddot{q}_{\alpha}}{N^2}+\frac{2}{3}g_Kq_{\alpha}^{-\frac{1}{3}}
 -2g_{\Lambda}q_{\alpha} -\frac{2}{3}g_{2}q_{\alpha}^{-\frac{5}{3}}-2g_{3}q_{\alpha}^{-3}\Bigr] \delta q_{\alpha} -
\frac{{\cal V}}{N}\dot{q}_{\alpha}\, \delta q_{\alpha}\mid^{t_f=1}_{t_i=0} + \delta S_B\,.   \label{variation}
\end{split}
\end{align}
To obtain the equation of motion for the scale factor $q_{\alpha}$, 
\begin{align}
\frac{\ddot{q}_{\alpha}}{N^2} = -\frac{2}{3}g_Kq_{\alpha}^{-\frac{1}{3}}
 +2g_{\Lambda}q_{\alpha} +\frac{2}{3}g_{2}q_{\alpha}^{-\frac{5}{3}}+ 2g_{3}q_{\alpha}^{-3}\,, \label{eomq}
\end{align} 
we have several choices of boundary conditions for $q_{\alpha}$ (Dirichlet, Neumann, and Robin ones)  as well as the specific boundary terms $S_B$ localized on the hypersurfaces at $t_{i,f}=0,1$.

When the boundary conditions are specified and we have solved the equation of motion \eqref{eomq}, we can evaluate the gravitational propagator under the semi-classical analysis as
\begin{equation}\label{G-Path-integral}
G\left[\{q_{\alpha}(t_f)\};\{q_{\alpha}(t_i)\}\right]   
 = \int \mathrm{d}N \int\mathcal{D}\mathfrak{q}  \, e^{iS\left[\mathfrak{q},N\right] / \hbar}
 = \sqrt{\frac{i\, {\cal V}}{2\pi\hbar}}\int_{0,-\infty}^\infty  \frac{\mathrm{d} N}{N^{1/2}} 
\exp \left(\frac{iS_{\rm on-shell}[\{q_{\alpha}(t_i),q_{\alpha}(t_f)\}, N]}{\hbar}\right)\,,
\end{equation}
where $S_{\rm on-shell}[\{q_{\alpha}(t_i),q_{\alpha}(t_f)\}, N]$ is the on-shell action for the background geometries. Here, for simplicity, the variables are written assuming the Dirichlet boundary condition. For different boundary conditions, \textit{e.g}. Neumann and Robin boundary conditions, the variables of the path integral are changed. Although the above integration does not converge, the Picard-Lefschetz theory complexifies the variables themselves and provides a unique way to find convergent integration contours along the steepest descent paths which is called Lefschetz thimbles $\cal J_\sigma$. This theory proceeds with the above integral as, 
\begin{equation}\label{G-Path-integral-pl}
G\left[\{q_{\alpha}(t_f)\};\{q_{\alpha}(t_i)\}\right]  =  \sqrt{\frac{i\, {\cal V}}{2\pi\hbar}}\sum_\sigma n_\sigma \int_{\cal J_\sigma} \frac{\mathrm{d} N}{N^{1/2}} 
\exp \left(\frac{iS_{\rm on-shell}[\{q_{\alpha}(t_i),q_{\alpha}(t_f)\}, N]}{\hbar}\right)\,,
\end{equation}
where $n_\sigma$ is the intersection number $\braket{\cal K_\sigma,\mathcal{R}}$ between the steepest ascent path $\cal K_\sigma$ and the original contour $\mathcal{R} $. By utilizing the Picard-Lefschetz theory~\cite{Witten:2010cx}, the integration over $N$ in equation~\eqref{G-Path-integral} of the propagator can be evaluated by identifying the relevant saddle points and Lefschetz thimbles $\cal J_\sigma$ in the complex $N$-plane.

For simplicity, we shall take the small universe limit $q_{\alpha}\lesssim (g_3/g_K)^{3/8}$, \textit{i.e}. the $z=3$ anisotropic scaling limit and assume that the higher-order term $g_{3}$ is dominated. In this approximation (we simply set $g_{\Lambda}=0$ and $g_2=0$), from the \ac{hl} action we derive the equations of motion,
\begin{align}
\frac{\ddot{q}_{\alpha}}{N^2}-\frac{2g_3}{q^{3}_{\alpha}}=0,
\end{align}
whose general solution is 
\begin{equation}\label{eq:hl-solution}
q_{\alpha}(t)=\pm \frac{\sqrt{b_{\alpha 1}^2
(b_{\alpha 2}+t)^2+2g_3N^2}}{\sqrt{b_{\alpha 1}}}.    
\end{equation}
Physically, $q_{\alpha}(t)$ expresses the squared scale factor and should be positive for the Lorentzian direction or real axis. Thus, we can neglect the negative sign. By using them we have the following on-shell action, 
\begin{equation}\label{eq:on-shell0}
S_{\rm on-shell}[\{q_{\alpha}(t)\}, N]
=\sum_{\alpha}{\cal V} \left\{
\sqrt{2g_{3}} \left(\tan ^{-1}\left[\frac{b_{\alpha 1} (b_{\alpha 2}+1)}{\sqrt{2g_{3}} N}\right]-\tan ^{-1}\left[\frac{b_{\alpha 1}b_{\alpha 2}}{\sqrt{2g_{3}} N}\right]\right)-\frac{b_{\alpha 1}}{2 N}\right\}\,,
\end{equation}
where we have set $S_B=0$. For the Neumann and Robin boundary conditions, we need to add appropriate boundary terms to the \ac{hl} action as shown later.

Up to this point we have not yet performed the path integral over the lapse. Therefore, the classical solution \eqref{eq:hl-solution} (after imposing proper boundary conditions) and the on-shell action \eqref{eq:on-shell0} (after adding proper boundary terms) are applicable to both projectable and non-projectable theories, and still contain the contribution from ``dark matter as integration constant''~\cite{Mukohyama:2009mz,Mukohyama:2009tp} in each local universe.

Once the path integral over the lapse is performed, however, the projectable and non-projectable theories deviate from each other. In the non-projectable theory, each local universe is equipped with an independent lapse function and, as a result, the integration over all lapse functions under the saddle point approximation forces the ``dark matter as integration constant'' component in each local universe to vanish. In the non-projectable theory, therefore the no-boundary wave function of the global universe is simply the direct product of wave functions of each local universe. On the other hand, in the projectable theory, the path integral over the common lapse function in the saddle point approximation does not force the ``dark matter as integration constant'' component in each local universe to vanish but leads to (anti) correlation between the ``dark matter as integration constant'' components in different local universes. After summing over all possible values of the ``dark matter as integration constant'' components that are consistent with the global Hamiltonian constraint, the no-boundary wave function of the global universe in the projectable theory inevitably contains entanglement between different local universes. In the next section, we perform the path integral over the lapse, focusing on the projectable theory.

\section{Hartle-Hawking no-boundary proposal}
\label{sec:no-boundary-proposal}

In this section, we shall discuss the no-boundary proposal based on the Lorentzian path integral, considering various boundary conditions and associated boundary terms for the background spacetime manifold. The Hartle-Hawking no-boundary proposal suggests that the wave function of the universe is defined by a path integral over all compact Euclidean geometries, with just a $3$-dimensional boundary on which the $3$-dimensional induced metric is required to agree with the argument of the wave function.
However, in its conventional and modernistic interpretation, the no-boundary path integral initiates from a universe with zero size and extends to one of finite size~\cite{Halliwell:1988ik}. Within the framework of Lorentzian quantum cosmology, the Dirichlet boundary condition is frequently utilized. We first address the Dirichlet boundary condition, followed by discussions on the Neumann and Robin boundary conditions for both initial and final hypersurfaces.

\subsection{Dirichlet boundary condition for initial and final hypersurfaces}

We consider the \emph{Dirichlet} boundary condition in the Lorentzian path integral, where we fix the value of the squared scale factor at the two endpoints, $q_{\alpha}\mid^{t_{f}=1}_{t_{i}=0}$. If we do not add any boundary term to the \ac{hl} action, $S_B=0$, then the Dirichlet boundary conditions for the squared scale factor $q_{\alpha}$ at $t_{i,f}=0,1$ are compatible with the variation of the action~\eqref{variation} as 
\begin{align}
 \ q_{\alpha}\mid^{t_{f}=1}_{t_{i}=0} = \textrm{fixed}\ 
 \Longrightarrow 
 -\frac{{\cal V}}{N}\dot{q}_{\alpha}\, \delta q_{\alpha}\mid^{t_{f}=1}_{t_{i}=0} = 0\,. 
 \label{Dirichlet}
\end{align}
We note that in the Einstein-Hilbert action of \ac{gr}, we have to impose the well-known Gibbons-Hawking-York (GHY) term~\cite{York:1972sj, Gibbons:1976ue}, $\int d^3x \sqrt{\gamma}K$ for the Dirichlet boundary conditions where $\gamma$ denotes the determinant of the induced metric on the boundary. However, by definition, there is no need to introduce such boundary terms in the \ac{hl} action.

We fix the squared scale factor $q_{\alpha}(t)$ on the initial and final hypersurfaces at $t_{i,f}=0,1$, and impose the following Dirichlet boundary conditions, 
\begin{equation}\label{eq:Dirichlet-boundary}
q_{\alpha}(t_{i}=0)=q_{\alpha i}, \quad q_{\alpha}(t_{f}=1)=q_{\alpha f},
\end{equation}
on the solution~\eqref{eq:hl-solution}. In this way, we can fix the integration constants as
\begin{equation}
b_{\alpha 1}=q_{\alpha i}^2+q_{\alpha f}^2 \mp 2 \sqrt{q_{\alpha i}^2q_{\alpha f}^2-2g_{3}N^2}, 
\quad b_{\alpha 2}=-\frac{q_{\alpha i}^2\mp \sqrt{q_{\alpha i}^2q_{\alpha f}^2-2g_{3}N^2}}{q_{\alpha i}^2+q_{\alpha f}^2\mp 2\sqrt{q_{\alpha i}^2q_{\alpha f}^2
-2g_{3} N^2}}.
\end{equation}
By using the above values of the integration constants, we have the on-shell action for the Dirichlet boundary conditions, 
\begin{align}\label{eq:on-shell-Dirichlet}
\begin{split}
S_{\rm on-shell}[\{q_{\alpha i},q_{\alpha f}\}, N]&
=\sum_{\alpha}{\cal V}\Biggl\{
\mp \sqrt{2g_{3}} \biggl(\tan ^{-1}\left[\frac{\mp q_{\alpha i}^2+\sqrt{q_{\alpha i}^2q_{\alpha f}^2-2g_{3}N^2}}{\sqrt{2g_{3}} N}\right]\\
&+\tan ^{-1}\left[\frac{\mp q_{\alpha f}^2+\sqrt{q_{\alpha i}^2q_{\alpha f}^2-2g_{3}N^2}}{\sqrt{2g_{3}} N}\right]\biggr)-\frac{q_{\alpha i}^2+q_{\alpha f}^2\mp 2 \sqrt{q_{\alpha i}^2q_{\alpha f}^2-2g_{3}N^2}}{2 N}\Biggr\}\,. 
\end{split}
\end{align}
Given that the global universe consists of individual local universes, the wave functions of these local universes influence each other and then, the critical points of the lapse function, which dictate the behavior of the overall wave function, are affected by the behavior of each individual local universe.

First, we consider the case with only one local universe $q_{1}(t)$. In this case, the critical points $N_c$ are given by setting the derivative of the on-shell action with respect to the lapse to zero, $\mathrm{d} S_{\rm on-shell} [q_{1 i},q_{1 f}, N]/\mathrm{d} N =0$, as
\begin{equation}
N_c=c_{1}\frac{i \left(q_0^2-q_{1 f}^2\right)}{2 \sqrt{2 g_3}}\,, 
\end{equation}
with $c_{1} \in \{-1 , +1\}$, where we have set $\{q_{1 i}\}=q_0$. We note that this result applies to each local universe in the non-projectable version as well.

\begin{figure}[t]
        \subfigure[$q_{1f}=q_{2f}=1$]{%
		\includegraphics[clip, width=0.24\columnwidth]
		{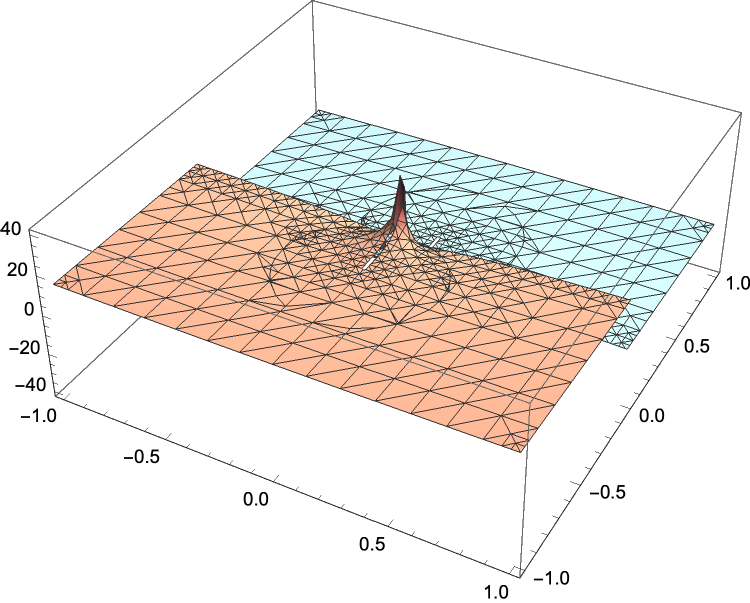}}%
	\subfigure[$q_{1f}=q_{2f}=1$]{%
		\includegraphics[clip, width=0.24\columnwidth]
		{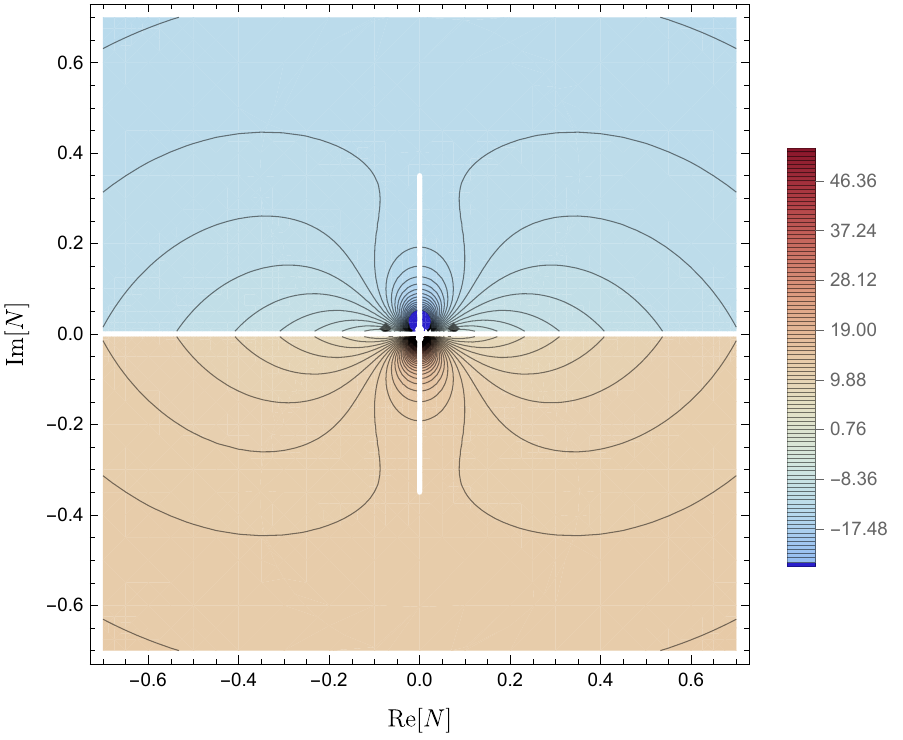}}
  \subfigure[$q_{1f}=1$ and $q_{2f}=5$]{%
		\includegraphics[clip, width=0.24\columnwidth]
		{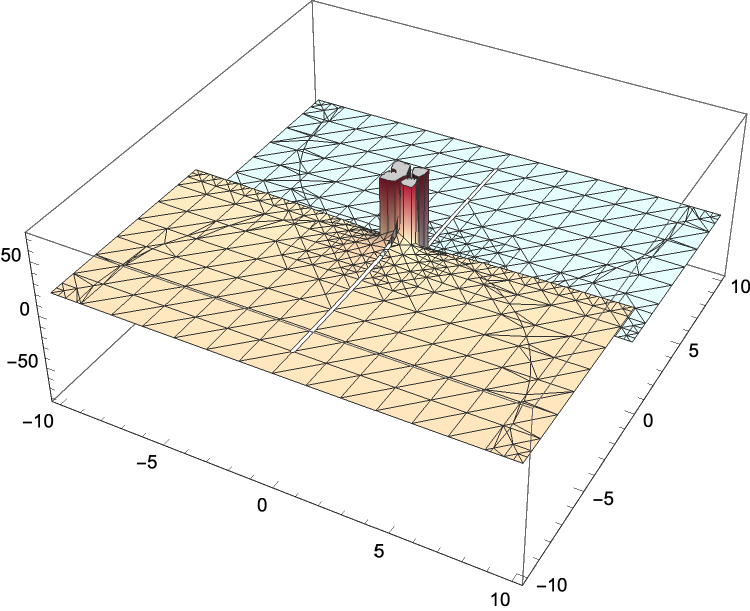}}%
	\subfigure[$q_{1f}=1$ and $q_{2f}=5$]{%
		\includegraphics[clip, width=0.24\columnwidth]
		{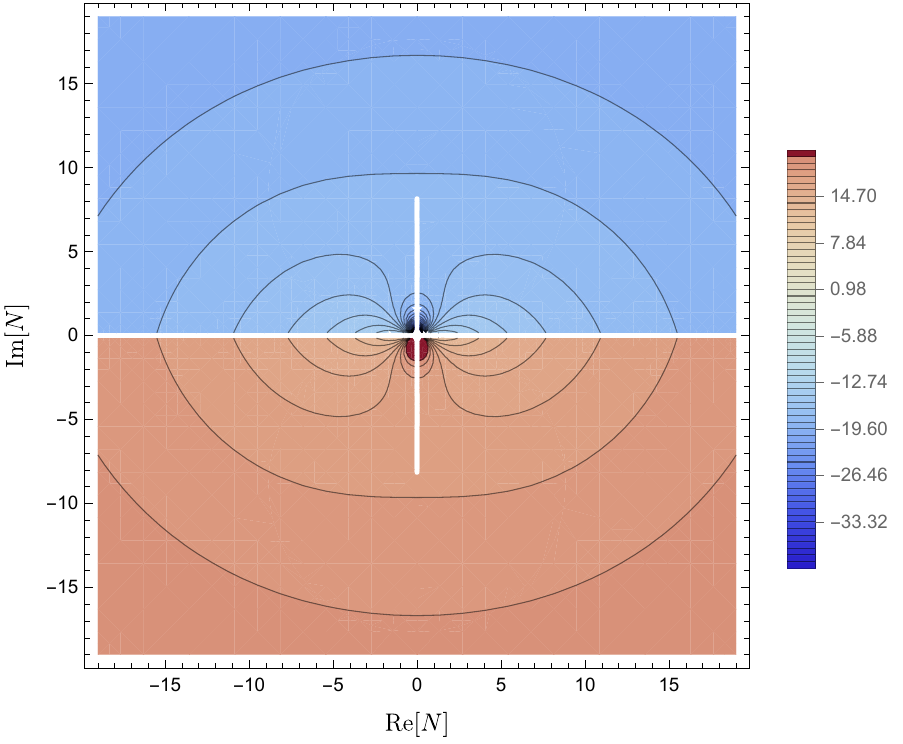}}
\subfigure[$q_{1f}=q_{2f}=1$]{%
		\includegraphics[clip, width=0.24\columnwidth]
		{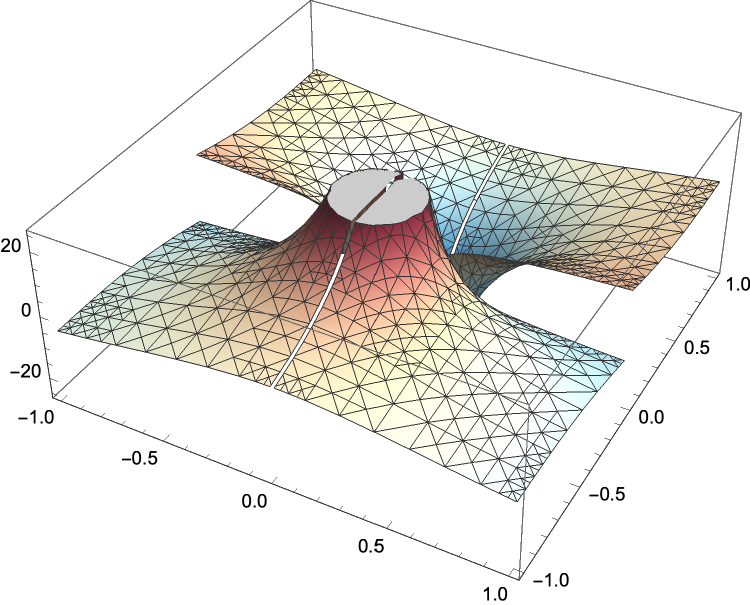}}%
	\subfigure[$q_{1f}=q_{2f}=1$]{%
		\includegraphics[clip, width=0.24\columnwidth]
		{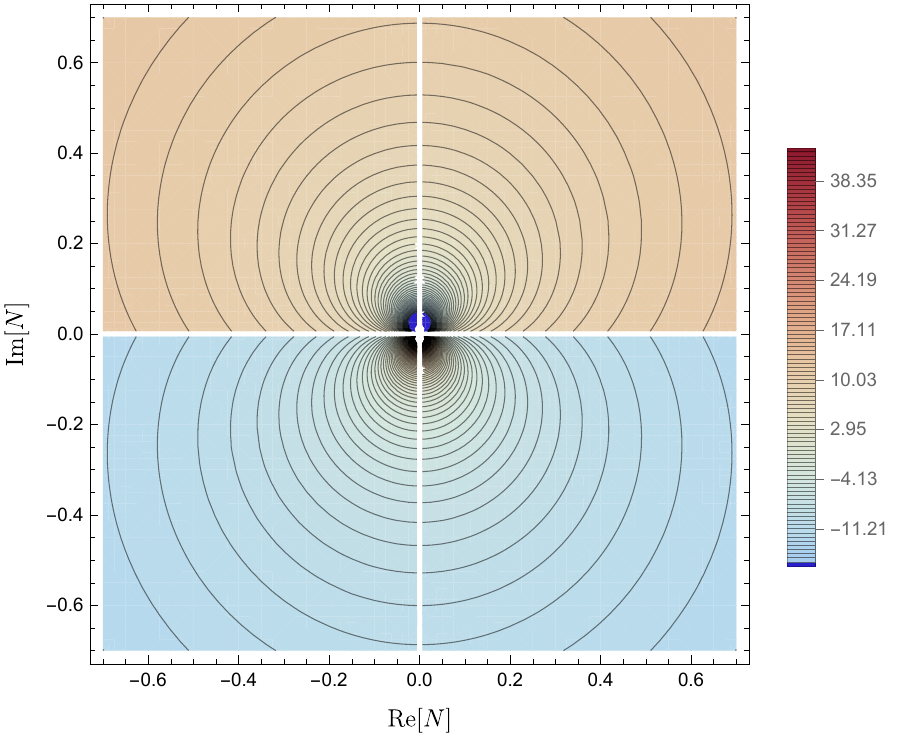}}
  \subfigure[$q_{1f}=1$ and $q_{2f}=5$]{%
		\includegraphics[clip, width=0.24\columnwidth]
		{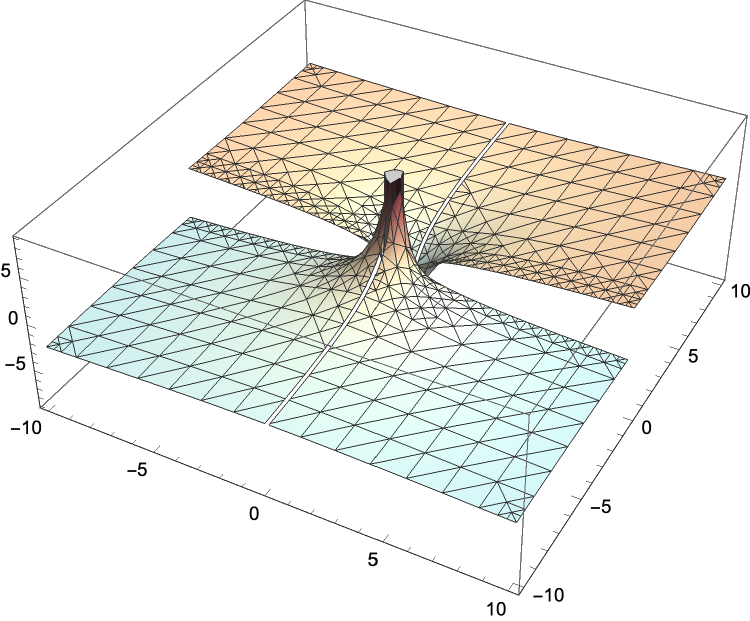}}%
	\subfigure[$q_{1f}=1$ and $q_{2f}=5$]{%
		\includegraphics[clip, width=0.24\columnwidth]
		{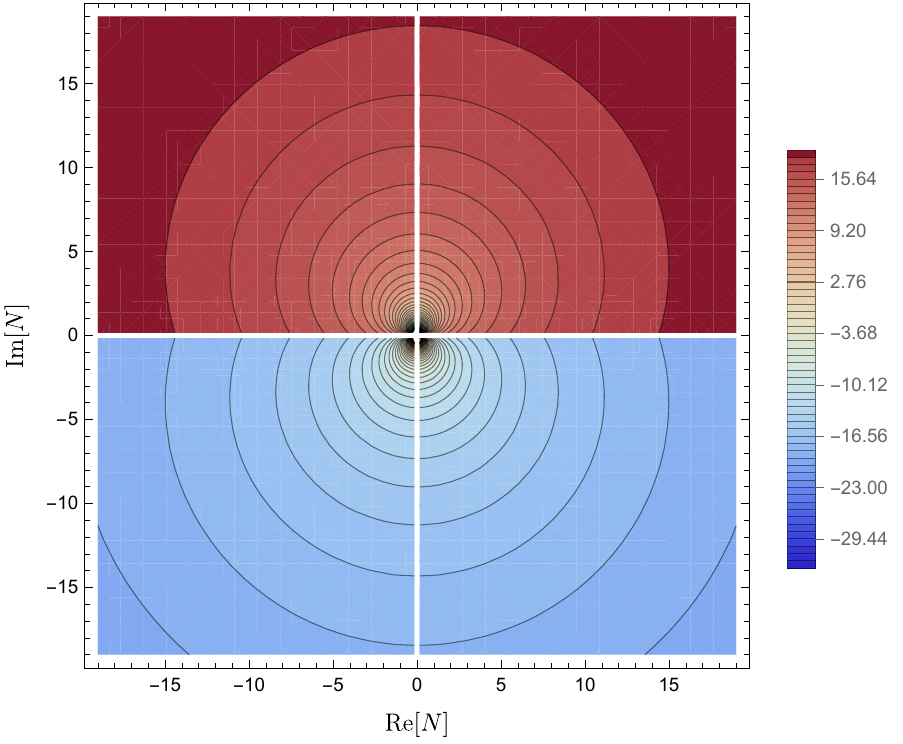}}
 \caption{We plot $\textrm{Re}\left[iS_{\rm on-shell}[N]\right]$ for the on-shell action of~\eqref{eq:on-shell-Dirichlet} with $\mp$ signs in the complex lapse plane. We fix $q_{0}=0.01$, $g_3=1$, ${\cal V}=1$ and take $-$ and $+$ signs of the on-shell action~\eqref{eq:on-shell-Dirichlet} in the top and bottom figures, respectively. The critical points are given by $N_c=\pm 0.353518 i, \ \pm 4.59616i$ for $q_{1f}=q_{2f}=1$, $q_{1f}=1$ and $q_{2f}=5$.  White lines express the branch cut on the complex $N$-plane. Here, in the top figures, where the $-$ sign is chosen for the on-shell action, the complex integral can be performed through the critical point by integrating along the imaginary axis avoiding the branch cut. On the other hand, in the bottom figures, which corresponds to the $+$ sign of the on-shell action, the complex $N$-plane is divided by branch cuts, and the Lefschetz thimble method can not be applied since these do not have pertinent saddle points.}
\label{fig:Picard-Lefschetz1}
\end{figure}

Next, we consider the case with a global universe consisting of two local universes $\{q_{1}(t),q_{2}(t)\}$. In this case the critical points $N_c$ given by $\mathrm{d} S_{\rm on-shell} [q_{1 i},q_{1 f},q_{2 i},q_{2 f}, N]/\mathrm{d} N =0$ are
\begin{equation}\label{eq:Dirichlet-Dirichlet-saddle-points}
N_c=c_{1}\frac{i\sqrt{16q_0^8 +8(q_{1f}^4-6 q_{2f}^2 q_{1f}^2+q_{2f}^4) q_0^4+(q_{1f}^2+q_{2f}^2)^4}}{4 \sqrt{2g_3}(2 q_0^2+q_{1f}^2+q_{2f}^2)}, 
\end{equation}
where $c_{1} \in \{-1 , +1\}$ and we have set $q_{\alpha i}=q_0$ for simplicity. We plot $\textrm{Re}\left[iS_{\rm on-shell}[N]\right]$ for the two local universes with the on-shell action~\eqref{eq:on-shell-Dirichlet} over the complex plane in Fig.~\ref{fig:Picard-Lefschetz1}. As with \ac{gr}, by approximating the critical points, the tunneling wave function can be found with the critical points with $\textrm{Im}[N] > 0$ where the lapse $N$ integration contour runs along the steepest descent paths and passes the critical points. On the other hand, the no-boundary wave function is given by the critical points with $\textrm{Im}[N] < 0$. The positive value of the critical points $c_1=+1$, corresponds to $\textrm{Im}[N] > 0$, while the negative value $c_1=-1$, corresponds to $\textrm{Im}[N] < 0$. Naively, the lapse $N$ integration contours to pass the no-boundary critical point correspond to the steepest ascent paths, so we can not obtain the no-boundary wave function. The issue of no-boundary wave function using Lorentzian path integral methods has been extensively discussed in Refs.~\cite{Feldbrugge:2017kzv,DiazDorronsoro:2017hti}. In the projectable \ac{hl} gravity with the Dirichlet boundary conditions, the Lorentzian path integral results in a tunneling wave function similar to \ac{gr}. However, we note that the analysis of the \ac{hl} wave function using the Lefschetz thimble method based on the Picard-Lefschetz theory becomes challenging in this case. We can only perform the complex integral through the tunneling critical point by integrating along the imaginary axis avoiding the branch cut.

Now, let us consider the quantum creation of the universe from nothing, where the Dirichlet boundary condition is imposed for the vanishing size of the scale factor at the initial time, i.e. the limit $q_0=\epsilon\to 0$, and leads to the coefficients,
\begin{equation}
\lim_{\epsilon \to 0} b_{\alpha 1}=q_{\alpha f}^2\mp 2 i\sqrt{2g_{3}} N\,,
\quad \lim_{\epsilon \to 0}
b_{\alpha 2}=\frac{N}{-2 N\mp \frac{iq_{\alpha f}^2}{\sqrt{2g_{3}}}}\,.
\end{equation}
In this case, we have the following on-shell actions,
\begin{align}\label{eq:on-shell-Dirichlet-divergence}
\lim_{\epsilon \to 0} S_{\rm on-shell}[\{\epsilon,q_{\alpha f}\}, N]&=
\sum_{\alpha}{\cal V}\Biggl\{
\sqrt{2g_{3}} \left(\mp i \tanh^{-1}\left(1\pm \frac{i q_{\alpha f}^2}{\sqrt{2g_{3}} N}\right)
\mp i\tanh^{-1}(1)\pm i\right) -\frac{q_{\alpha f}^2}{2 N}\Biggr\}\,,
\end{align}
where $\tanh^{-1}(\pm i)=\pm i\infty$, and the above on-shell action diverges. On the other hand, under the $z=3$ anisotropic scaling, the marginal and irrelevant boundary terms localized on the initial hypersurface at $t_{i}=0$ cannot contain more than one time derivative, more than three spatial derivatives or any mixed derivatives, meaning that they are not divergent and that the divergent bulk contributions to the on-shell action cannot be canceled by boundary contributions. Therefore, for the Dirichlet boundary conditions with $\{q_{\alpha i}\}=0$ the semi-classical approximation should break down. In this situation, it is expected that the \ac{rg} running of coupling constants should be significant. In the literature, it has been argued that the \ac{uv} fixed point of the \ac{rg} flow is characterized by the limit $\lambda \to \infty $, where the theory is weakly coupled~\cite{Gumrukcuoglu:2011xg,Radkovski:2023cew}. Also, in the \ac{uv}, $g_3$ plays an important role. It is therefore natural to consider the \ac{rg} flow of $\lambda$, $g_{3}$. One can show that under a specific phenomenological parameterization of the \ac{rg} flow of $\lambda$, $g_{3}$, the divergences of the on-shell action~\eqref{eq:on-shell-Dirichlet-divergence} introduced by the higher dimensional operators can be eliminated. We provide a detailed discussion in Appendix~\ref{sec:appendix1}.

Under the above consideration, let us neglect the contribution from the on-shell action at $q_{\alpha i}\to 0$. The derivative of the on-shell actions reads,
\begin{align}
\lim_{\epsilon \to 0}\frac{\mathrm{d} 
S_{\rm on-shell}[\{\epsilon,q_{\alpha f}\}, N]}{\mathrm{d} N}=
\sum_{\alpha=1}^n{\cal V}\Biggl\{\frac{q_{\alpha f}^2\mp 
2\sqrt{-2g_{3}N^2}}{2 N^2}\Biggr\}=
{\cal V}\Biggl\{\frac{\sum_{\alpha=1}^n q_{\alpha f}^2\mp 
2n\sqrt{-2g_{3}N^2}}{2 N^2}\Biggr\}\,, 
\end{align}
where we have counted the number ($\alpha=1,\dots, n$) of the local universes. Therefore, we find the critical points $N_c$ given by $\mathrm{d} S_{\rm on-shell}[\{q_{\alpha f}\}, N]/\mathrm{d} N =0$,
\begin{equation}
\lim_{\epsilon \to 0} N_c=\frac{ic_{1}\sum_{\alpha=1}^n q_{\alpha f}^2}{2n\sqrt{2g_3}}\,,
\end{equation}
with $c_{1} \in \{-1 , +1\}$. For a global universe consisting of a couple of local universes, this agrees with \eqref{eq:Dirichlet-Dirichlet-saddle-points} in the limit $q_0=\epsilon\to 0$. The corresponding critical points are imaginary which means that the global universe satisfying the global Hamiltonian constraint behaves in a quantum way as long as $g_3>0$. In the limit $\epsilon \to 0$ and approximating the critical points the \ac{hl} wave function of the global universe only including two local universes $\{q_{1}(t),q_{2}(t)\}$ can be given by, 
\begin{align}
\begin{split}
\Psi(q_{1f},q_{2f}) & \simeq \sqrt{\frac{2{\cal V}\sqrt{2g_{3}}}{c_1\pi\hbar\left(q_{1f}^2+q_{2f}^2\right)} }
\exp \Biggl(\frac{c_1{\cal V}\sqrt{2g_{3}}}{\hbar}\Biggl\{
 \tanh ^{-1}\left(\frac{\mp \left(q_{1f}^2+q_{2f}^2\right)+4 q_{1f}^2}{c_1 \left(q_{1f}^2+q_{2f}^2\right)}\right)\\ & \qquad\qquad +\tanh ^{-1}\left(\frac{\mp \left(q_{1f}^2+q_{2f}^2\right)+4 q_{2f}^2}{c_1 \left(q_{1f}^2+q_{2f}^2\right)}\right) 
-2 \left(\mp 1+1\right) \Biggr\}\Biggr)\,.
\end{split}
\end{align}
From Fig.~\ref{fig:Picard-Lefschetz1} we can only perform the complex integral through the tunneling critical point with $c_1=1$ by integrating along the imaginary axis avoiding the branch cut and approximately get 
\begin{equation}
\Psi(q_{1f},q_{2f})=\sqrt{\frac{2{\cal V}\sqrt{2g_{3}}}
{\pi\hbar\left(q_{1f}^2+q_{2f}^2\right)} }\left(\frac{4 q_{1f}^2 q_{2f}^2}{\left(q_{1f}^2-q_{2f}^2\right)^2}\right)^{\frac{{\cal V}\sqrt{2 g_3}}{\hbar }}\,,
\end{equation}
where we used $\tanh ^{-1}(x)= \log\left({1+x}/{1-x}\right)^{\frac{1}{2}}$.

\subsection{Robin boundary condition for initial and final hypersurfaces}

As previously mentioned, the Dirichlet boundary condition can be employed to reduce the size of the universe to zero on the initial hypersurface, embodying the concept of quantum creation of the universe from nothing. This boundary condition, however, introduces challenges in the stability of perturbations in the framework of quantum cosmology. These perturbative stability concerns have been highlighted in recent studies~\cite{Feldbrugge:2017fcc,Feldbrugge:2017mbc,Matsui:2022lfj}, shedding light on the need for alternative approaches, \textit{e.g}. considering non-trivial boundary conditions in gravitational theories. As previously discussed, the on-shell action in \ac{hl} gravity diverges due to the higher-dimensional terms when the universe's size is set to zero, and therefore, we need to consider the \ac{rg} flow of $g_{3}$ and $\lambda$. In the following, we will explore non-trivial boundary conditions, \textit{i.e}. \emph{Neumann} and \emph{Robin} boundary conditions within the framework of projectable \ac{hl} gravity, and show that the above issues can be easily avoided by such non-trivial boundary conditions.

The Neumann boundary condition fixes the derivative of the field at the two endpoints, $\dot{q}_{\alpha}\mid^{t_{f}=1}_{t_{i}=0}$~\cite{Krishnan:2016mcj}. In the \ac{hl} gravity we have to impose the GHY term $S_B$ to the \ac{hl} action, 
\begin{equation}
S_B=\frac{1}{\zeta}\sum_{\alpha} \int_{\Sigma_{\alpha}}d^3x \sqrt{\gamma}K= \sum_{\alpha}\frac{{\cal V}}{N}q_{\alpha}\dot{q}_{\alpha}\mid^{t_{f}=1}_{t_{i}=0}\,,
\end{equation}
where $\zeta=\frac{3}{2(3\lambda-1)}$ is a constant. Here, the constants of the boundary terms exhibit \ac{rg} dependence. The variation of the boundary term removes the term in $\delta q_{\alpha}$ and replaces it with a term in $\delta \dot{q}_{\alpha}$ only,
\begin{align}
\frac{1}{N}q_{\alpha}\,\delta\dot{q}_{\alpha}\mid^{t_{f}=1}_{t_{i}=0} = 0\ 
\Longrightarrow \ \frac{1}{N}\dot{q}_{\alpha}\mid^{t_{f}=1}_{t_{i}=0} = \textrm{fixed} \,. 
\end{align}
Although the Neumann boundary conditions can specify the momentum of early and late time scale factors, it has been pointed out that the Euclidean initial momentum is crucial for the success of the no-boundary proposal. Imposing the Euclidean (imaginary) Neumann boundary conditions on the initial hypersurface, the perturbative instability can be avoided in GR~\cite{DiTucci:2019dji,DiTucci:2019bui}. Unfortunately, it is found that the \ac{hl} action does not provide the analytical and explicit solutions $q_{\alpha}(t)$ or the on-shell action. For this technical reason, we do not adopt the Neumann boundary conditions in this paper.

Next, we will study the Robin boundary condition. The Robin boundary condition is a mixed boundary condition that includes both the field value and its derivative, combining aspects of both the Dirichlet and Neumann boundary conditions. It introduces a balance between the field value and its derivative at the boundary, providing a more flexible approach to the boundary issues. Although not as common as the Dirichlet and Neumann boundary conditions, Robin boundary conditions have proved their usefulness in gravitational problems as perturbation problems of Euclidean gravity~\cite{Witten:2018lgb}. On the other hand, it is not straightforward to impose Robin boundary terms covariantly in \ac{gr}. Ensuring that the boundary terms are defined in a covariant way is crucial for maintaining the validity of the gravitational theory~\cite{York:1972sj,Gibbons:1976ue,Krishnan:2017bte}~\footnote{For example, to impose the Robin boundary condition we can impose the boundary term on the action~\cite{Krishnan:2017bte},
\begin{equation*}
S_B=\frac{1}{\zeta}\sum_{\alpha} \int_{\Sigma_{\alpha}} d^3x \sqrt{\gamma}K + 
\frac{1}{\varsigma}\sum_{\alpha} \int_{\Sigma_{\alpha}} d^3x \sqrt{\gamma}\,, 
\end{equation*}
where $\zeta,\varsigma$ are constants. From our metric, the variation of the action leads to 
\begin{equation*}
\frac{q_{\alpha}}{N}\delta\dot{q}_{\alpha}\mid^{t_{f}=1}_{t_{i}=0}  + 
\frac{2q_{\alpha}}{\tilde{\varsigma}}\delta q_{\alpha}\mid^{t_{f}=1}_{t_{i}=0}\, =
q_{\alpha}\,\delta\left(\frac{\dot{q}_{\alpha}}{N} + 
\frac{2q_{\alpha}}{\tilde{\varsigma}} \right)\mid^{t_{f}=1}_{t_{i}=0}=0 \,, 
\end{equation*}
where $\tilde{\varsigma}=\frac{4(3\lambda-1)}{3}\varsigma$. Taking the specific boundary conditions we have 
\begin{equation*}
\frac{1}{N}\frac{\dot{q}_{\alpha}}{q_{\alpha}}\mid^{t_{f}=1}_{t_{i}=0}=-\frac{2}{\tilde{\varsigma}}=
\frac{3}{2(1-3\lambda)\varsigma}\,, 
\end{equation*}
which corresponds to the Hubble expansion rate. However, the constant $\varsigma, \tilde{\varsigma}$ pertains solely to the covariant boundary term and is not associated with individual local universes. As a result, these boundary terms do not allow one to use the Hubble expansion rate as the boundary data or the argument of the wave function of the universe. Thus, we do not adopt this boundary term.}.

In order to impose the Robin boundary condition that allows one to fix the Hubble expansion rate for the initial and final configurations, we shall utilize the following boundary term, 
\begin{equation}\label{eq:robin-boundary-term}
S_B=\frac{1}{2\, \zeta}\sum_{\alpha} \int_{\Sigma_{\alpha}}d^3x \sqrt{\gamma}K=
\sum_{\alpha}\frac{{\cal V}}{2N}q_{\alpha}\dot{q}_{\alpha}\mid^{t_{f}=1}_{t_{i}=0}= \sum_{\alpha}\frac{{\cal V}}{2N}q_{\alpha}^2
\frac{\dot{q}_{\alpha}}{q_{\alpha}}\mid^{t_{f}=1}_{t_{i}=0}\,, 
\end{equation}
and the variation of the boundary term leads to, 
\begin{align}
\delta S_B&=\sum_{\alpha}\frac{{\cal V}}{N}q_{\alpha}\delta q_{\alpha} \frac{\dot{q}_{\alpha}}{q_{\alpha}}\mid^{t_{f}=1}_{t_{i}=0} + \sum_{\alpha}\frac{{\cal V}}{2N}q_{\alpha}^2\delta \left(\frac{\dot{q}_{\alpha}}{q_{\alpha}} \right)
\mid^{t_{f}=1}_{t_{i}=0}\notag \\
&= \sum_{\alpha}\frac{{\cal V}}{N}\dot{q}_{\alpha}\delta q_{\alpha}\mid^{t_{f}=1}_{t_{i}=0} + \sum_{\alpha}\frac{{\cal V}}{2N}q_{\alpha}^2\delta \left(\frac{\dot{q}_{\alpha}}{q_{\alpha}} \right)
\mid^{t_{f}=1}_{t_{i}=0}\,.
\end{align}
The variation of the \ac{hl} action including the above boundary term removes the term in $\delta q_{\alpha}$ and replaces it with a term in $\delta \dot{q}_{\alpha}$ only,
\begin{align}
\frac{1}{N}q_{\alpha}^2\delta \left(\frac{\dot{q}_{\alpha}}{q_{\alpha}} \right) = 0\ 
\Longrightarrow \ \frac{1}{N}\frac{\dot{q}_{\alpha}}{q_{\alpha}}\mid^{t_{f}=1}_{t_{i}=0} = \textrm{fixed} \,. 
\end{align}
Thus, we obtain the boundary condition on the Hubble expansion rate at the initial and final hypersurface, ${H}_{\alpha}=\frac{1}{Na_{\alpha}}\frac{\mathrm{d} a_{\alpha}}{\mathrm{d} t}=\frac{2}{3N}\frac{\dot{q}_{\alpha}}{q_{\alpha}}$, and the Robin boundary condition seems to be very convenient and physical. The method with the above boundary term to introduce the Robin boundary conditions is not known. Therefore, for comparison, we discuss the case of \ac{gr} in Appendix~\ref{sec:appendix2}.

Hereafter, we will assume such Robin boundary conditions on the initial hypersurface and the final hypersurface for the scale factor,
\begin{equation}\label{eq:robin-boundary}
\frac{1}{2N}
\frac{\dot{q}_{\alpha}(t_{i}=0)}{q_{\alpha}(t_{i}=0)}
=\frac{1}{\xi_{\alpha i}}, \quad \frac{1}{2N}
\frac{\dot{q}_{\alpha}(t_{f}=1)}{q_{\alpha}(t_{f}=1)}
=\frac{1}{\xi_{\alpha f}}\,,
\end{equation}
where $\xi_{\alpha i},\xi_{\alpha f}$ are constants. By imposing these conditions on the solution~\eqref{eq:hl-solution}, we can fix the integration constants as
\begin{align}
b_{\alpha 1}&=\mp\frac{2 i \sqrt{g_3} N^{3/2} \left(\xi_{\alpha i}-\xi_{\alpha f}+4 N\right)}{\sqrt{\xi_{\alpha i} \xi_{\alpha f}^2-\xi_{\alpha i}^2 \xi_{\alpha f}+8 N^3+8 \xi_{\alpha i} N^2-8 \xi_{\alpha f} N^2+2 \xi _{\alpha i}^2 N+2 \xi_{\alpha f}^2 N-6 \xi_{\alpha i} \xi_{\alpha f} N}}\,, \label{eq:b_alpha1_Robin} \\
b_{\alpha 2}&=\frac{\xi_{\alpha f}-2 N}{\xi_{\alpha i}-\xi_{\alpha f}+4 N}\,.\label{eq:b_alpha2_Robin}
\end{align}

By substituting the solutions~\eqref{eq:hl-solution} with \eqref{eq:b_alpha1_Robin}-\eqref{eq:b_alpha2_Robin} to the action with the above boundary terms $S_B$~\eqref{eq:robin-boundary-term}, we can obtain the on-shell action for the Robin boundary condition~\eqref{eq:robin-boundary},
\begin{align}\label{eq:on-shell-Robin}
\begin{split}
S_{\rm on-shell}[\{\xi_{\alpha}\},N]&=
\sum_{\alpha=1}^n{\cal V}\Biggl\{i \sqrt{2 g_3} \Biggl(\tanh ^{-1}\left(\frac{\mp \sqrt{2 N} \left(2 N- \xi_{\alpha f}\right)}{\sqrt{\left(\xi_{\alpha i}+2 N\right) \left(2 N-\xi_{\alpha f}\right) \left(\xi_{\alpha i}-\xi_{\alpha f}+2 N\right)}}\right) \\
& \mp\tanh ^{-1}\left(\frac{\sqrt{2 N} \left(\xi_{\alpha i}+2 N\right)}{\sqrt{\left(\xi_{\alpha i}+2 N\right) \left(2 N-\xi_{\alpha f}\right) \left(\xi_{\alpha i}-\xi_{\alpha f}+2 N\right)}}\right)\Biggr)\Biggr\}\,,
\end{split}
\end{align}
where we have counted the number ($\alpha=1,\dots, n$) of the local universes. We note that the on-shell action does not diverge unless the initial or final Hubble ratio is taken to infinity and the no-boundary proposal does not necessarily require infinite Hubble expansion rate at the initial hypersurface. Thus, we can consider the finite initial and final Hubble expansion rates. The derivative of the on-shell action with respect to the lapse reads,
\begin{align}
\frac{\mathrm{d} S_{\rm on-shell}[\{\xi_{\alpha}\}, N]}{\mathrm{d} N}
=\sum_{\alpha=1}^n{\cal V}\left\{\mp\frac{i \sqrt{g_3} \left(\xi_{\alpha i}-\xi_{\alpha f}+4 N\right)}{\sqrt{N} \sqrt{\left(\xi_{\alpha i}+2 N\right) \left(2 N-\xi_{\alpha f}\right) \left(\xi_{\alpha i}-\xi_{\alpha f}+2 N\right)}}\right\}=0\,.
\end{align}

\begin{figure}[t]
        \subfigure[$\xi_{1f}=\xi_{2f}=10$]{%
		\includegraphics[clip, width=0.24\columnwidth]
		{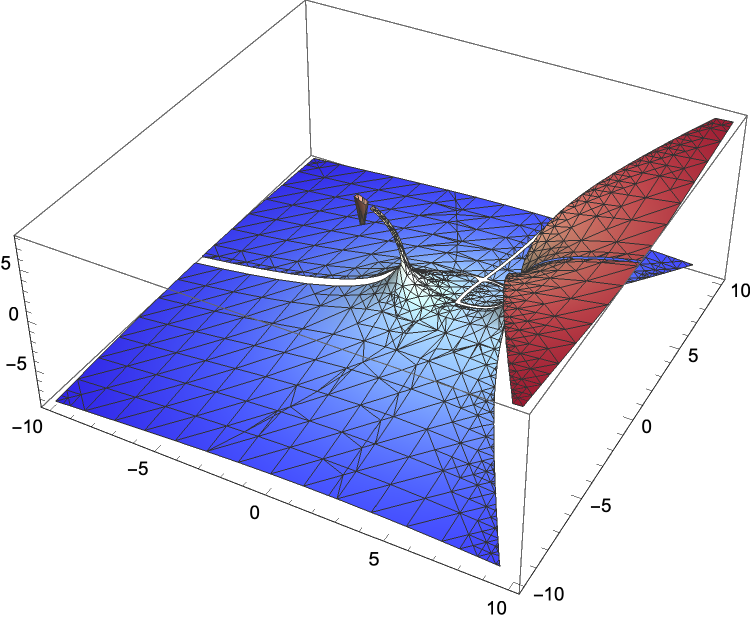}}%
	\subfigure[$\xi_{1f}=\xi_{2f}=10$]{%
		\includegraphics[clip, width=0.24\columnwidth]
		{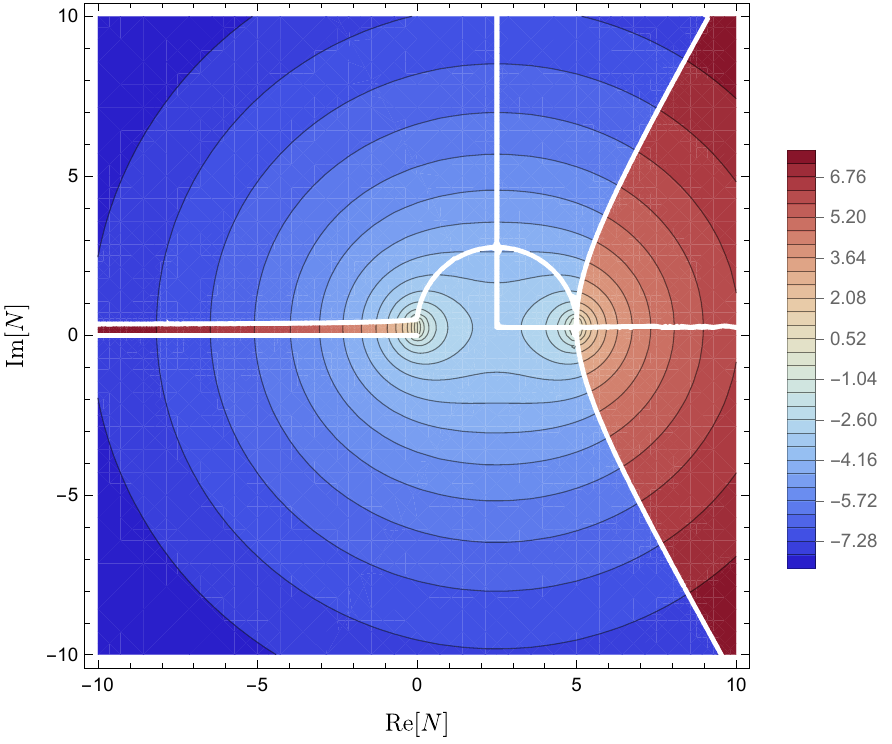}}
  \subfigure[$\xi_{1f}=10$ and $\xi_{2f}=50$]{%
		\includegraphics[clip, width=0.24\columnwidth]
		{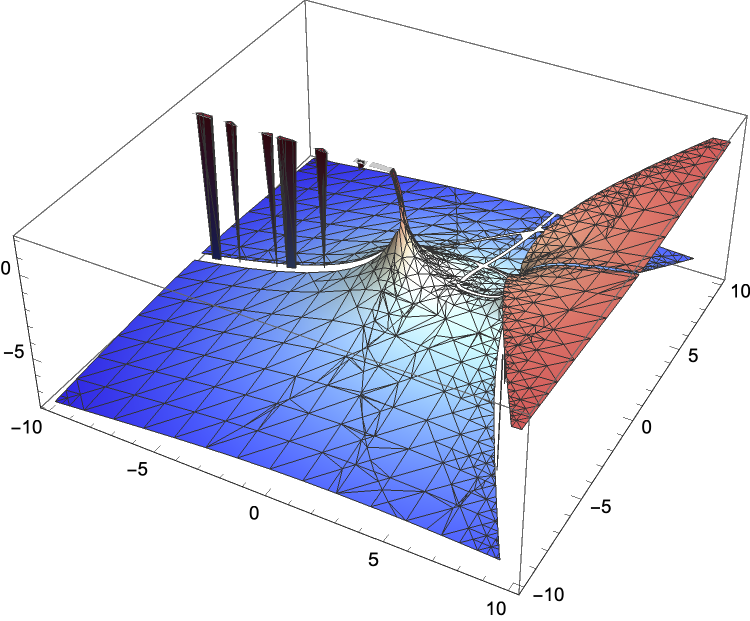}}%
	\subfigure[$\xi_{1f}=10$ and $\xi_{2f}=50$]{%
		\includegraphics[clip, width=0.24\columnwidth]
		{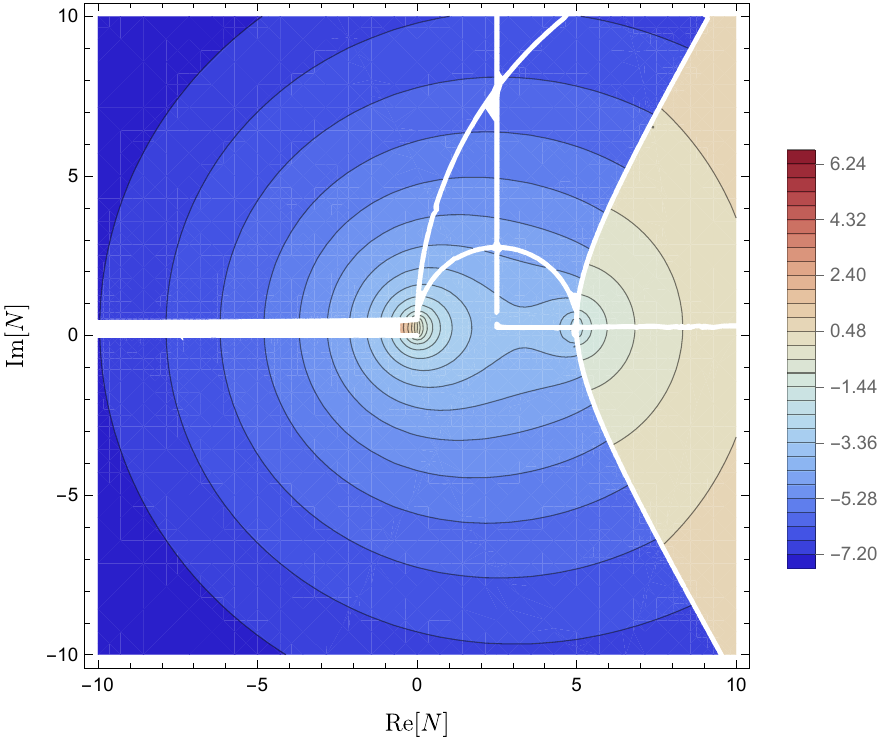}}
\subfigure[$\xi_{1f}=\xi_{2f}=10$]{%
		\includegraphics[clip, width=0.24\columnwidth]
		{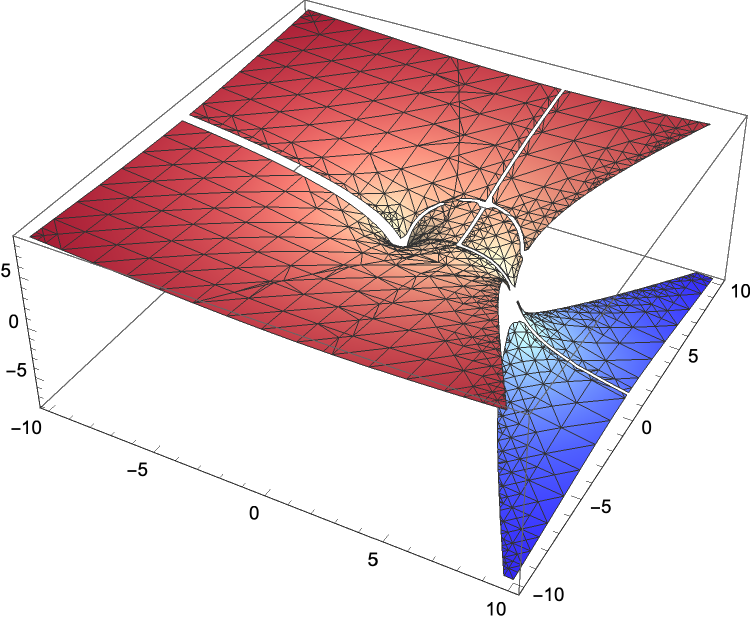}}%
	\subfigure[$\xi_{1f}=\xi_{2f}=10$]{%
		\includegraphics[clip, width=0.24\columnwidth]
		{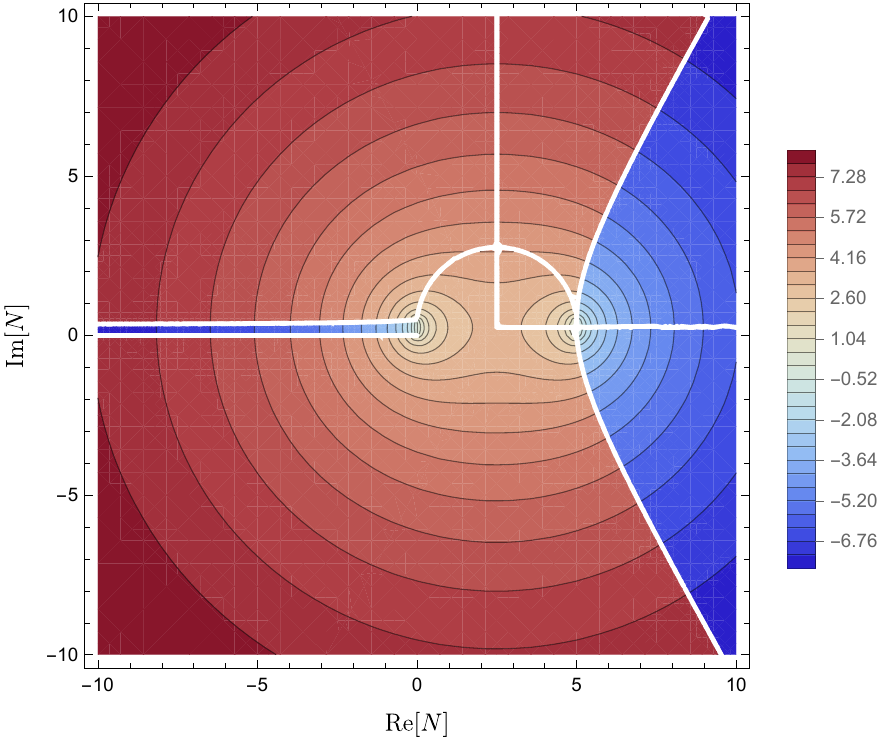}}
  \subfigure[$\xi_{1f}=10$ and $\xi_{2f}=50$]{%
		\includegraphics[clip, width=0.24\columnwidth]
		{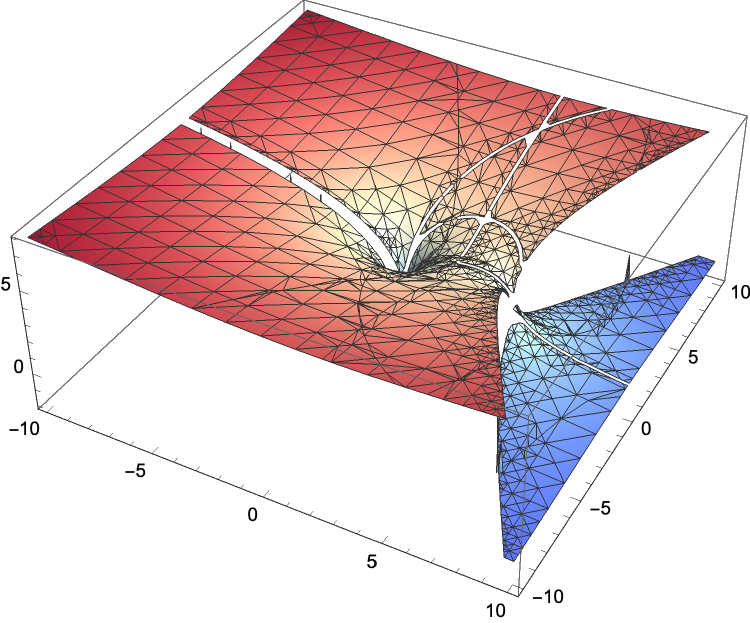}}%
	\subfigure[$\xi_{1f}=10$ and $\xi_{2f}=50$]{%
		\includegraphics[clip, width=0.24\columnwidth]
		{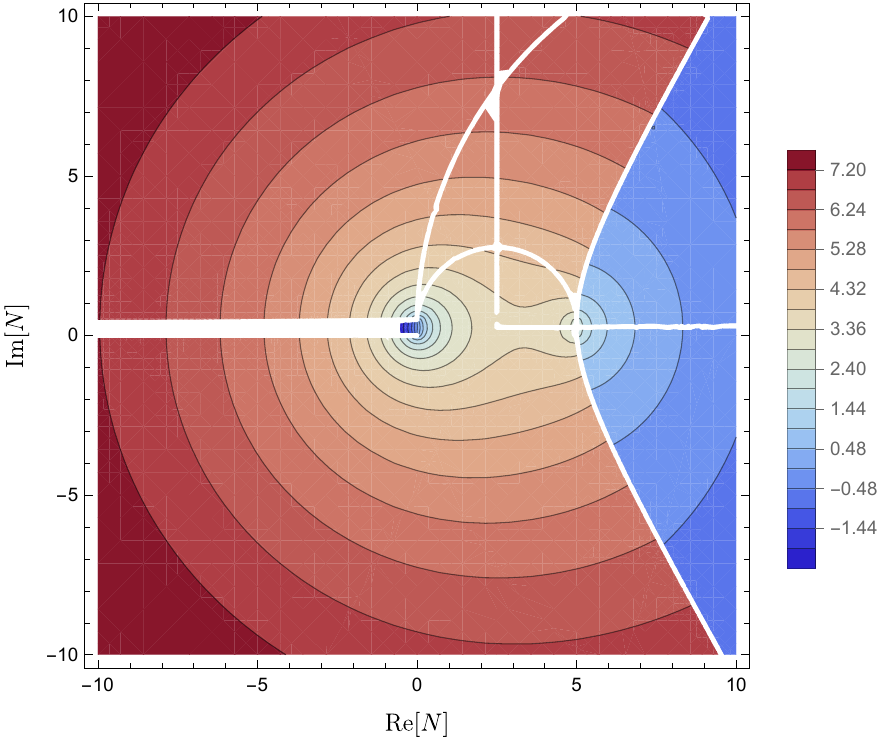}}
\caption{We plot $\textrm{Re}\left[iS_{\rm on-shell}[N]\right]$ for the on-shell action of~\eqref{eq:on-shell-Robin} with $\mp$ signs in the complex lapse plane. We fix $\xi_0=-i$, $g_3=1$, 
${\cal V}=1$ and take $-$ and $+$ signs of the on-shell action~\eqref{eq:on-shell-Robin} in the top and bottom figures, respectively. There are several branch cuts of the on-shell action for the Robin boundary condition, and each is derived from the square root $\sqrt{z}$ or inverse hyperbolic function $\tanh ^{-1}(z)$, which has branch cuts at $-1$ and $1$ on the real axis. In particular, in the top figures, where the on-shell action has the $-$ sign, we can perform the complex integral avoiding branch cuts. The complex path of the integral always corresponds to $\textrm{Im}[N] < 0$ and the no-boundary wave function is achieved.}
\label{fig:Picard-Lefschetz2}
\end{figure}

First, we consider the case with only one local universe $q_{1}(t)$. In this case, the critical points $N_c$ are given by setting the derivative of the on-shell action with respect to the lapse to zero, $\mathrm{d} S_{\rm on-shell} [\xi_{1 i},\xi_{1 f}, N]/\mathrm{d} N =0$, as 
\begin{equation}
N_c=\frac{\xi_{1 f}-\xi_{1 i}}{4} \,.
\end{equation}
Next, taking $\xi_{\alpha i}=0$ we can obtain  
\begin{align}
\frac{\mathrm{d} S_{\rm on-shell}[\{\xi_{\alpha}\}, N]}{\mathrm{d} N}
=\sum_{\alpha=1}^n{\cal V}\left\{\mp\frac{i \sqrt{g_3} \left(4 N-\xi_{\alpha f}\right)}{\sqrt{2} \sqrt{N} \sqrt{N \left(2 N-\xi_{\alpha f}\right)^2}}\right\}=0.
\end{align}

For the global universe only including two local universes $\{q_{1}(t),q_{2}(t)\}$, we find the critical points $N_c$, 
\begin{equation}
N_c=\frac{3 \xi_{1f}+3 \xi_{2f}+c_{1}\sqrt{9 \xi_{1f}^2-14 \xi_{2f} \xi_{1f}
+9 \xi_{2f}^2}}{16} \,,
\end{equation}
with $c_{1} \in \{-1 , +1\}$. For a real Hubble expansion rate at the final hypersurface, they are also real.

As already mentioned, in \ac{gr} it has been known that by imposing the imaginary Neumann boundary conditions on the initial hypersurface, the perturbative instabilities can be avoided and the no-boundary wave function can be obtained without mixing with tunneling wave function~\cite{DiTucci:2019dji,DiTucci:2019bui}. Hereafter, following this, in \ac{hl} gravity, we shall explore the Robin boundary condition with a particular imaginary Hubble expansion rate at the initial hypersurface. According to the Picard-Lefschetz theory, given a saddle point, the behavior of the integral can be exactly estimated. However, it is not always possible to evaluate the integral using the Lefschetz thimble method and in the on-shell action of the \ac{hl} gravity, in particular, with regard to Robin boundary conditions, the deformation of the integral path by the Lefschetz thimbles is non-trivial. Therefore, to obtain a meaningful result, we will consider the appropriate integration path in the complex plane by using the contour plots with specifying numerical values.

In Fig.~\ref{fig:Picard-Lefschetz2}, we plot $\textrm{Re}\left[iS_{\rm on-shell}[N]\right]$ for the two local universes with the on-shell action~\eqref{eq:on-shell-Robin} over the complex plane. To take the Robin boundary conditions with a particular imaginary Hubble expansion rate at the initial hypersurface, we fixed $\xi_0=-i$, $g_3=1$, ${\cal V}=1$ and take $-$ or $+$ sign of the on-shell action~\eqref{eq:on-shell-Robin} in the top and bottom figures. The corresponding and initial Robin boundary condition implies 
\begin{align}
    \frac{1}{2N}
\frac{\dot{q}_{\alpha}(t_{i}=0)}{q_{\alpha}(t_{i}=0)}
=\frac{1}{-i}=+i\,.
\end{align}
There are several branch cuts of the on-shell action~\eqref{eq:on-shell-Robin}, and each is derived from the square root $\sqrt{z}$ or inverse hyperbolic function $\tanh ^{-1}(z)$, which has branch cuts at $-1$ and $1$ on the real axis. In particular, in the on-shell action~\eqref{eq:on-shell-Robin} has $-$ signs, we can perform the complex integral with avoiding branch cuts. The complex path of the integral always corresponds to $\textrm{Im}[N] < 0$ and the Hartle-Hawking no-boundary wave function is realized. By taking the imaginary and initial Hubble ratio $\textrm{Im}[\xi_0] < 0$, we can derive the no-boundary wave function. Conversely, with $\textrm{Im}[\xi_0] > 0$, we obtain the tunneling wave function. However, the tunneling wave function is known to exhibit perturbative instability, as shown by~\cite{Feldbrugge:2017fcc,Feldbrugge:2017mbc}, and perhaps this $\textrm{Im}[\xi_0] > 0$ should not be taken.

\section{Discussions and conclusions}
\label{sec:discussions-conclusions}

In this work, we have studied the Hartle-Hawking no-boundary proposal within the framework of \ac{hl} gravity, mainly focusing on the projectable theory. The no-boundary proposal is a prominent hypothesis that describes the quantum creation of the universe from nothing, while the \ac{hl} gravity is a potential theory of quantum gravity that ensures renormalizability and unitarity, at least in the projectable version. We have adopted the modern formulation of the no-boundary proposal by using the Lorentzian path integral which enables us to describe the quantum evolution of the universe from nothing. For simplicity, we have focused on a global universe composed of a set of local universes each of which is homogeneous, isotropic, and closed. It has been found that the application of the no-boundary proposal to the \ac{hl} gravity is not inherently simple. In particular, due to the presence of the higher-order curvature operators in the \ac{hl} gravity, the on-shell action for the no-boundary proposal naively diverges for the Dirichlet boundary condition. We then discuss the possibility that the divergence of the on-shell action due to higher-dimensional operators may be ameliorated by taking into account the \ac{rg} flow of coupling constants $\lambda$ and $g_{3}$. For the Robin boundary condition, the on-shell action does not diverge as long as the initial or final Hubble ratio is taken to finite. For these boundary conditions, we have explored the complex $N$-integral to obtain the Ho\v{r}ava-Lifshitz wave function. For the Dirichlet boundary condition, we identify pertinent critical points and integration contour, and the complex integral can be performed through one critical point by integrating along the imaginary axis avoiding the branch cut. On the other hand, in the on-shell action~\eqref{eq:on-shell-Robin} under Robin boundary conditions, the deformation of the integral path by the Lefschetz thimbles is non-trivial. We have encountered several branch cuts of the on-shell \ac{hl} action originating from the square root $\sqrt{z}$ or inverse hyperbolic function $\tanh ^{-1}(z)$. For the Robin boundary condition with a positively imaginary Hubble rate at the initial hypersurface, we can perform the complex integral while avoiding branch cuts. The complex path of the integral that avoids branch cuts always corresponds to $\textrm{Im}[N] < 0$ and the Hartle-Hawking no-boundary wave function is realized. Although we have indeed demonstrated that no-boundary and tunneling wave functions can be formulated within the framework of \ac{hl} gravity, their formulations are conditional, as discussed in this paper. On the other hand, the DeWitt wave function is readily formulated and consistent with perturbation theory as shown in our previous works~\cite{Matsui:2021yte,Martens:2022dtd}. Our results shed light on the non-trivial relationship between the framework of quantum gravity theory and the wave function of the universe.

In projectable \ac{hl} gravity, the wave function of the global universe with multiple local universes inevitably contains entanglement between local universes. The origin of the entanglement is the global Hamiltonian constraint, which states that the total sum of the ``dark matter as integration constant''~\cite{Mukohyama:2009mz,Mukohyama:2009tp} should be zero. For example, for a global universe consisting of two local universes, if the energy density of the ``dark matter as integration constant'' in each local universe with the volume $V_{1,2}$ is $\rho_{{\rm DM}\, 1,2}$ then the global Hamiltonian constraint states that $\rho_{{\rm DM}\, 1}V_1+\rho_{{\rm DM}\, 2}V_2=0$, showing the anti-correlation between the amount of the ``dark matter as integration constant'', $\rho_{{\rm DM}\, 1,2}V_{1,2}$, in the two local universes. The no-boundary wave function of the global universe contains contributions from realizations of the global universe with all possible values of $\rho_{{\rm DM}\, 1,2}V_{1,2}$ with the complete anti-correlation between the two local universes. For this reason, the wave function of the global universe inevitably contains entanglement between the local universes.

On the other hand, in non-projectable \ac{hl} gravity, each local universe has its own local lapse function and thus its own local Hamiltonian constraint. The local Hamiltonian constraint sets the ``dark matter as integration constant'' to vanish at each point in each local universe. As a result, the no-boundary wave function of the global universe in non-projectable \ac{hl} gravity is simply the direct product of wave functions of each local universe, without any entanglement between local universes. Therefore, as far as we are interested in observables in one of the local universes in the no-boundary proposal of non-projectable \ac{hl} gravity, there is no difference between the case with only one local universe and that with multiple local universes. Hence, it suffices to consider the global universe consisting of one local universe and, as a result, the no-boundary wave function in non-projectable \ac{hl} gravity is obtained by simply restricting that in projectable \ac{hl} gravity to the case with the global universe consisting of only one local universe.

In the present paper, we have mainly focused on the application of the projectable \ac{hl} gravity to the Hartle-Hawking no-boundary proposal, restricting our consideration to the anisotropic scaling limit of $z=3$. 
Beyond the anisotropic scaling limit, the analysis of the no-boundary wave function up to the \ac{ir} limit as the usual scaling of $z=1$, is left as future work. 
If this wave function can be analytically evaluated up to the \ac{ir}, it might be possible to qualitatively assess the probabilistic predictions and quantum entanglement of the wave function of each local universe dominated by dark matter as an integral constant.
Furthermore, we have focused only on the wave function for the background and have not evaluated the perturbations. The treatment of perturbations in quantum cosmology is a notoriously difficult problem, and a thorough examination of the perturbation analysis in the wave function of this universe is necessary. However, the perturbation problem is expected to be resolved by the renormalizability of the \ac{hl} gravity or the Robin boundary condition.

\medskip
{\it Acknowledgments.}
The authors would like to thank Atsushi Naruko and Paul Martens for collaborating on the previous works~\cite{Matsui:2021yte,Martens:2022dtd,Matsui:2022lfj} on the subject. H.M. would like to thank Masazumi Honda for the helpful discussions. The work of H.M. was supported by JSPS KAKENHI Grant No. JP22KJ1782 and No. JP23K13100. The work of S.M.~was supported in part by the World Premier International Research Center Initiative (WPI), MEXT, Japan. S.M. is grateful for the hospitality of Perimeter Institute where part of this work was carried out. Research at Perimeter Institute is supported in part by the Government of Canada through the Department of Innovation, Science and Economic Development and by the Province of Ontario through the Ministry of Colleges and Universities.

\appendix

\section{UV salvation for on-shell divergences}
\label{sec:appendix1}

In this appendix, we shall demonstrate that the divergence of the on-shell action with the Dirichlet boundary condition~\eqref{eq:on-shell-Dirichlet-divergence} can in principle disappear once the \ac{rg} flow of $\lambda$ and $g_{3}$ is taken into account.

Since the divergence of the on-shell action stems from the higher spatial derivative terms, we set $\alpha_2=c_{\rm g}^2=\Lambda=0$ in \eqref{eq:HLaction-FLRW}. For the same reason, we ignore the ``dark matter as integration constant'', which means that it suffices to consider the global universe consisting of only one local universe. Then the \ac{hl} action becomes 
\begin{equation} 
        S\left[\{a_{\alpha}\},N\right] =  V\!\int_{t_i}^{t_f} \mathrm{d} t\, 
        Na^3 \left[ -\frac{3(3\lambda-1)}{2}\left(\frac{\dot{a}}{Na}\right)^{2}
              + \frac{\alpha_3}{a^6} \right]\,, 
\end{equation}
where we have supposed $\alpha_3>0$ for classical stability. As stated in the main text, since the semi-classical approximation breaks down, we need to take into account the \ac{rg} running of coupling constants. However, since the \ac{rg} properties of \ac{hl} gravity have not yet been understood well and we do not know how to implement the \ac{rg} running of coupling constants into the computation of the path integral, we adopt a phenomenological approach where the coupling constants in the action are promoted to functions of $a$ as
\begin{equation}
 \frac{3(3\lambda-1)}{2} \rightarrow f(a)\,, \quad
  \alpha_3 \rightarrow h(a)\,, 
\end{equation}
so that
\begin{equation} 
        S\left[\{a_{\alpha}\},N\right] =  V\!\int_{t_i}^{t_f} N \mathrm{d} t\, L\,, \quad 
        L = a^3 \left[ -f(a)\left(\frac{\dot{a}}{Na}\right)^{2}
             + \frac{h(a)}{a^6} \right]\,. 
\end{equation}
We then suppose that $f(a)$ and $h(a)$ have the following parametrized forms near $a=0$.
\begin{equation}
  f(a) \propto a^{-2\alpha}\,, \quad h(a) \propto a^{2\beta}\,, \quad (\mbox{near } a=0)\,,
\end{equation} 
where $\alpha>0$ so that $\lambda$ runs towards $+\infty$ in the UV, i.e. in the limit $a\to +0$, as suggested by \cite{Gumrukcuoglu:2011xg,Radkovski:2023cew}.

The equation of motion for $a$ then leads 
\begin{equation}
 a^{2-\alpha-\beta}\partial_{\tau} a = const.\,, \quad (\mbox{near } a=0)\,, 
\end{equation}
where we have introduced the proper time $\tau$ as usual by $d\tau = N dt$. By assuming $3-\alpha-\beta>0$ and properly shifting the origin of $\tau$ so that $a\to +0$ as $\tau\to +0$, we obtain
\begin{equation}
 a \propto \tau^{\frac{1}{3-\alpha-\beta}}\,, \quad (\mbox{near } \tau=0)\,.
\end{equation}
By substituting this to $L$, we obtain
\begin{equation}
 L = -\frac{2h(a)}{a^3} \propto \tau^{-1+\frac{\beta-\alpha}{3-\alpha-\beta}}\,, \quad (\mbox{near } \tau=0)\,.
\end{equation}
Therefore, the on-shell action does not exhibit the UV divergence if 
\begin{equation}
 0 < \alpha < \beta < 3-\alpha\,.
\end{equation}

\section{Robin boundary condition in general relativity}
\label{sec:appendix2}

As discussed in the body of the paper, it is not easy to impose Robin boundary terms covariantly in gravitational theories.  Within the \ac{gr} framework, the simplest boundary terms that implement Robin boundary conditions were introduced in the literature~\cite{Krishnan:2017bte}. Subsequent research, particularly the work~\cite{DiTucci:2019bui} explored the Lorentzian path integral to the Robin boundary conditions for the no-boundary proposal.  Our current work presents a different Robin boundary term in the \ac{hl} gravity. In the following appendix, we will consider the implications of the no-boundary proposal with this Robin boundary term in \ac{gr} for comparison.

First, we assume a closed Friedmann-Lema\^{i}tre-Robertson-Walker (FLRW) universe, 
\begin{equation}
 N = N(t)\,, \quad N^i = 0\,, \quad g_{ij} = a^2(t) \left[\Omega_{ij} ({\bf x}) \right]\,.
\end{equation}
With these assumptions 3+1 dimensional action in \ac{gr} reads, 
\begin{align}
S\left[a,N\right] &=  2\pi^2 \int \mathrm{d} t\, 
\left({Na^3}\right) \left[-3\left(\frac{\dot{a}}{Na}\right)^{2}
              + \frac{3}{a^2}-\Lambda \right]\,,
\end{align}
where we take the Planck mass unit with $M_{\rm Pl}=1/\sqrt{8\pi G}=1$. To simplify our analysis in \ac{gr}, we introduce the new time coordinate $\tau$ that is related to $t$ by  
\begin{equation}\label{new-time-gr}
 a(t) \mathrm{d}t=\mathrm{d}\tau \,.
\end{equation}
For the closed FLRW universe, the gravitational propagator can be written as,
\begin{equation}\label{gr-propagator}
 G\left[\mathfrak{q}(\tau_f);\mathfrak{q}(\tau_i)\right] = \int \mathrm{d}N
 \int
\mathcal{D}\mathfrak{q}  \, e^{iS\left[\mathfrak{q},N\right] / \hbar} \,,
 \end{equation}  
where the lapse $N$ has been gauge-fixed to be time-independent by the BFV procedure, $\mathfrak{q}(\tau)=a^2(\tau)$. We then proceed with the lapse integral and the path integral over all configurations of the scale factor with the given initial $(\tau_i=0)$ and final $(\tau_f=1)$ values.

To evaluate the above path integral we proceed with the semi-classical analysis, and utilize the classical solutions of the action where we add possible boundary contributions $S_B$ localized on the hypersurfaces at $\tau_{i,f}=0,1$,
\begin{align}
S\left[\mathfrak{q},N\right]
=2\pi^2 \int_{\tau_i=0}^{\tau_f=1} Nd\tau 
 \Biggl[-\frac{3}{4N^2}{\mathfrak{q}'}^2
 +3-\Lambda \mathfrak{q}\Biggr]+S_B ,
\end{align}
where $'$ means the derivative of $\tau$.
Hereafter, we proceed with the variation of the action and derive the equation of motion. Since the action $S\left[\mathfrak{q},N\right]$ depends on $\mathfrak{q}$, ${\mathfrak{q}'}$, the variation of the action is given by 
\begin{align}
\begin{split}
\delta S\left[\mathfrak{q},N\right] = 2\pi^2 \int_{\tau_i=0}^{\tau_f=1} Nd\tau \, \Bigl[\frac{3}{2N^2}{\mathfrak{q}''}-\Lambda\Bigr] \delta \mathfrak{q} - 2\pi^2
\left(\frac{3}{2N}{\mathfrak{q}'}\, \delta \mathfrak{q}\mid^{\tau_f=1}_{\tau_i=0} \right)
+ \delta S_B\,.   
\end{split}
\end{align}
Thus, we derive the equation of motion for the squared scale factor $\mathfrak{q}$, 
\begin{align}
\frac{3}{2N^2}{\mathfrak{q}''}=\Lambda\,, \label{eq:gr-eom}
\end{align} 
whose general solution is 
\begin{equation}\label{eq:gr-solution}
\mathfrak{q}(t)=b_2 t+b_1+\frac{3}{4} \Lambda  N^2 t^2.    
\end{equation}
Now, we have to impose the boundary conditions for $\mathfrak{q}(t)$ as well as to add specific boundary terms $S_B$ localized on the hypersurfaces at $\tau_{i,f}=0,1$. As discussed in Section~\ref{sec:no-boundary-proposal} we consider the Robin boundary condition at the initial and final hypersurface, and shall utilize the following boundary term, 
\begin{equation}
S_B=-\frac{2}{3}\int d^3x \sqrt{\gamma}K=
2\pi^2
\left(\frac{1}{N}{\mathfrak{q}'} \mathfrak{q}\mid^{\tau_f=1}_{\tau_i=0} \right)= 2\pi^2
\left(\frac{1}{N} \mathfrak{q}^{3/2}\frac{{\mathfrak{q}'}}{\mathfrak{q}^{1/2}}\mid^{\tau_f=1}_{\tau_i=0} \right)\,, 
\end{equation}
where $K=-\frac{3}{N}\frac{\dot{a}}{a}$ and 
the variation leads to 
\begin{align}
\delta S_B&=2\pi^2
\left(\frac{3}{2N}{\mathfrak{q}'}\delta \mathfrak{q}\mid^{\tau_f=1}_{\tau_i=0} \right) 
+ 2\pi^2
\left(\frac{1}{N} \mathfrak{q}^{3/2} \delta
\left(\frac{{\mathfrak{q}'}}{\mathfrak{q}^{1/2}}\right)
\mid^{\tau_f=1}_{\tau_i=0} \right)\,.
\end{align}
The variation of the \ac{hl} action including the above boundary term results in a term with $\delta ({{\mathfrak{q}'}}/{\mathfrak{q}^{1/2}})$ only,
\begin{align}
\frac{1}{N} \mathfrak{q}^{3/2} \delta
\left(\frac{{\mathfrak{q}'}}{\mathfrak{q}^{1/2}}\right) = 0\ 
\Longrightarrow \ \frac{1}{N}\frac{{\mathfrak{q}'}}{\mathfrak{q}^{1/2}}\mid^{\tau_f=1}_{\tau_i=0} = \textrm{fixed} \,. 
\end{align}
Thus, we obtain the Robin boundary condition at the initial and final hypersurface. This boundary condition precisely corresponds to the Hubble expansion rate given by ${H}=\frac{1}{Na}\frac{\mathrm{d} a}{\mathrm{d} t}=\frac{1}{2N}\frac{{\mathfrak{q}'}}{\mathfrak{q}^{1/2}}$, as discussed in Section~\ref{sec:no-boundary-proposal},
and suggests its relevance in considering the evolution of the universe from a physical perspective.

Hereafter, we will assume such robin boundary conditions on the initial hypersurface and the final hypersurface,
\begin{equation}\label{eq:hubble-boundary-gr}
\frac{1}{N}
\frac{\mathfrak{q}(\tau_{i}=0)'}{\mathfrak{q}(\tau_{i}=0)^{1/2}}
=\frac{1}{\xi_{i}}, \quad \frac{1}{N}
\frac{\mathfrak{q}(\tau_{f}=1)'}{\mathfrak{q}(\tau_{f}=1)^{1/2}}
=\frac{1}{\xi_{f}}\,,
\end{equation}
where $\xi_{i},\xi_{f}$ are constants. By imposing these conditions on the solution~\eqref{eq:gr-solution}, we can get the coefficients,
\begin{align}
b_{1}&=\frac{\xi_i^2 (3 \Lambda  \xi _f^2-1) 
\left(2c_1 \sqrt{N^4 (3 \Lambda  \xi _f^2-1) 
\left(3 \Lambda  \xi _i^2-1\right)}+N^2 \left(3 \Lambda 
(\xi _f^2+\xi _i^2)-2\right)\right)}{4(\xi _f^2-\xi _i^2)^2}\,, \\
b_{2}&=\frac{-c_1\sqrt{N^4(3 \Lambda  \xi _f^2-1)(3 \Lambda  \xi _i^2-1)}
-3\Lambda N^2\xi_f^2+N^2}{2(\xi _f^2-\xi _i^2)}\,, 
\end{align}
with $c_{1} \in \{-1 , +1\}$.

When the boundary conditions are specified and we have solved the equation of motion \eqref{eq:gr-eom}, we can evaluate the gravitational propagator \eqref{gr-propagator} under the semi-classical analysis and obtain the following expression by using the Picard-Lefschetz theory, 
\begin{align}\label{G-Path-integral-gr}
G\left[\xi_{i},\xi_{f}\right]  
= \sum_\sigma n_\sigma \int_{\cal J_\sigma} \mathrm{d}N
P(\hbar,N)
\exp \left(\frac{iS_{\rm on-shell}[\xi_{i},\xi_{f}, N]}{\hbar}\right)\,,  
\end{align}
where $P(\hbar,N)$ is the prefactor and 
$S_{\rm on-shell}[\xi_{i},\xi_{f}, N]$ 
is the on-shell action for the background geometries,
\begin{align}\label{eq:on-shell-gr}
\begin{split}
&S_{\rm on-shell}[\xi_{i},\xi_{f},N]=\frac{\pi ^2 N}{8(\xi _f^2-\xi _i^2)^2}
\Biggl(c_1 \left(\Lambda  \xi _f^2 \left(12 \Lambda  \xi _i^2+1\right)+\Lambda  \xi _i^2-2\right) \sqrt{N^4(3 \Lambda  \xi _f^2-1) \left(3 \Lambda  \xi _i^2-1\right)}+\xi _f^4 \left(18 \Lambda ^3 N^2 \xi _i^2-\Lambda ^2 N^2+48\right)\\
&+\xi _f^2 \left(18 \Lambda ^3 N^2 \xi _i^4-\xi _i^2 \left(13 \Lambda ^2 N^2+96\right)-\Lambda  N^2\right)+N^2 \left(3 \Lambda  \xi _f^2-1\right) \left(3 \Lambda  \xi _i^2-1\right)+\xi _i^4 \left(48-\Lambda ^2 N^2\right)-\Lambda  N^2 \xi _i^2+N^2\Biggr)\,.
\end{split}
\end{align}
Utilizing the Picard-Lefschetz theory~\cite{Witten:2010cx}, the integration over $N$ of the propagator can be evaluated by identifying the relevant critical points and Lefschetz thimbles $\cal J_\sigma$ in the complex $N$-plane.
The derivative of the on-shell action reads,
\begin{align}
\begin{split}
&\frac{\mathrm{d} S_{\rm on-shell}[\xi_{i},\xi_{f}, N]}{\mathrm{d} N}=
\frac{3 \pi ^2}{8(\xi _f^2-\xi _i^2)^2}
\Biggl(
c_1 \left(\Lambda  \xi _f^2 \left(12 \Lambda  \xi _i^2+1\right)+\Lambda  \xi _i^2-2\right) \sqrt{N^4(3 \Lambda  \xi _f^2-1) \left(3 \Lambda  \xi _i^2-1\right)}\\
&+\xi _f^4 \left(18 \Lambda ^3 N^2 \xi_i^2-\Lambda ^2 N^2+16\right)+2 \xi _f^2 \left(9 \Lambda ^3 N^2 \xi _i^4-2 \xi _i^2 \left(\Lambda ^2 N^2+8\right)-2 \Lambda  N^2\right)+\xi _i^4 \left(16-\Lambda ^2 N^2\right)-4 \Lambda  N^2 \xi _i^2+2 N^2\Biggr)\,.
\end{split}
\end{align}
Therefore, we founds the critical points $N_c$ given by $\mathrm{d} S_{\rm on-shell}[\xi_{i},\xi_{f}, N]/\mathrm{d} N =0$,
\begin{equation}
N_c = 4c_{2} \sqrt{\frac{A_1 +c_{3}A_2 - 2}{A_3}}\,,
\end{equation}
where 
\begin{align}
\begin{split}
A_1 &= \Lambda  \left(\Lambda  \xi _f^4 \left(1-18 \Lambda  \xi _i^2\right)+\xi _f^2 \left(2 \Lambda  \xi _i^2 \left(2-9 \Lambda  \xi _i^2\right)+4\right)+\Lambda  \xi _i^4+4 \xi _i^2\right)\,, \\
A_2 &= \sqrt{(3 \Lambda  \xi _f^2-1)(3 \Lambda  \xi _i^2-1) \left(\Lambda  \xi _f^2 \left(12 \Lambda  \xi _i^2+1\right)+\Lambda  \xi _i^2-2\right)^2}\,, \\
A_3 &= \Lambda ^2 \left(\Lambda  \left(\Lambda  \xi _f^4 \left(1-18 \Lambda  \xi _i^2\right)^2+\xi _f^2 \left(11-\Lambda  \xi _i^2 \left(36 \Lambda  \xi _i^2+71\right)\right)+\Lambda  \xi _i^4+11 \xi _i^2\right)-1\right)\,,
\end{split}
\end{align}
with $c_{2,3} \in \{-1 , +1\}$. When limiting $\xi_i \to 0$
and positing that the universe originates from the initial singularity,
the critical points $N_c$ reduces, 
\begin{equation}
N_c = 
\frac{4c_{2} \xi _f^2}{\sqrt{\Lambda ^2 \xi _f^4\pm \left(4 \Lambda  \xi _f^2-2+\sqrt{(\Lambda\xi _f^2-2)^2(1-3 \Lambda  \xi _f^2)}\right)}}\,.
\end{equation}
For example, assuming $0<\xi _f<\sqrt{\frac{(5\sqrt{5}-11)}{2\Lambda}}$, the corresponding critical points $N_c$ through which the Lefschetz thimbles $\cal J_\sigma$ pass are real. This implies that the universe behaves classically. We have confirmed these facts numerically. For the Dirichlet boundary condition applied to a closed universe in \ac{gr}, it is important to note that the critical points are complex. Conversely, under the Robin boundary condition, if the boundary condition is real, the corresponding critical points are also real. This result is consistent with \ac{hl} gravity, as shown in Section~\ref{sec:no-boundary-proposal}.

\bibliographystyle{JHEP}
\bibliography{reference}

\end{document}